\documentclass[preprint,amsmath,amssymb,aps]{revtex4-1}

\usepackage{graphicx}
\usepackage{dcolumn}
\usepackage{bm}
\usepackage{epsfig}
\usepackage{epstopdf}
\usepackage{subfig}
\usepackage{caption}
\usepackage{enumerate}
\usepackage{comment}
\usepackage{physics}
\usepackage{amsmath}
\usepackage{slashed}
\usepackage{hyperref}
\usepackage [autostyle, english = american]{csquotes}
\MakeOuterQuote{"}
\newcommand{\Lim}[1]{\raisebox{0.5ex}{\scalebox{1.0}{$\displaystyle \lim_{#1}\;$}}}

\begin{document}

\title{Equivalence of Light-Front and Covariant Quantum Electrodynamics at One Loop Level and  the form of the  Gauge Boson Propagator}

\author{Deepesh Bhamre}
 \email{deepesh.bhamre@physics.mu.ac.in}
\author{Anuradha Misra}
 \email{misra@physics.mu.ac.in}
\affiliation{Department of Physics, University of Mumbai\\
 Santacruz (East), Mumbai-400098, India}

\date{\today}

\begin{abstract}
We consider the three fundamental one loop Feynman diagrams of QED viz. vertex correction, fermion self-energy and vacuum polarization in the light-front gauge and discuss the equivalence of their standard covariant expressions with the light-front expressions obtained using light-cone time-ordered Hamiltonian perturbation theory. Although this issue has been considered  by us and others previously, our emphasis in this article is on  addressing  the  ambiguity regarding the correct form of the gauge boson propagator to be used in the light-front gauge. We generalize our earlier results and show, using an alternative method called the Asymptotic Method,  how integrating over the light-front energy consistently  in  the covariant expression of each of the three one loop corrections leads to the propagating as well as the instantaneous diagrams of the light-front theory. In doing so, we re-establish the necessity of using the correct form of gauge boson propagator.

\begin{description}

\item[Keywords]
{Light-Front, QED}
\item[PACS numbers]
{11.15.Bt,12.20.-m}

\end{description}

\end{abstract}

\maketitle

\section{Introduction}\label{sec:intro}

In the recent past, the issue of equivalence of the covariant theory and light-front time-ordered Hamiltonian perturbation theory (LFTOPT) has attracted a lot of attention \citep{bakkerdewitt, ligterinkbakker, schoonthesis, schoonderbakker9857,schoonderbakker9858, paston}. Issue of equivalence in theories involving scalars and spin-$\frac{1}{2}$ particles has been discussed in Ref.\citep{ligterinkbakker}, whereas Refs.\cite{schoonthesis, schoonderbakker9857, schoonderbakker9858} deal with equivalence in Yukawa theory. Ref.\citep{paston} has considered the issue of equivalence in QED in (1+1) dimensions. The equivalence of light-front QED (LFQED) in light-front (LF) gauge and conventional QED in Coulomb gauge has been addressed in Ref.\cite{rohrlich} within the framework of Feynman-Dyson-Schwinger theory.

\par

The recent interest in this topic is related to the issue of renormalization of LF theories \citep{mustaki, ligterinkbakker2}. In light-front calculations, one  uses Hamiltonian perturbation theory and starting with the LF Hamiltonian $P^{-}$ and, using   the Heitler method of old fashioned time-ordered perturbation theory, arrives at the expressions of LF field theory. Mustaki  {\it et al.} have obtained the LF expressions  for one loop graphs of LFQED using this method in Ref.\citep{mustaki} and the same method has been used by authors of Ref.\citep{zhangharindranath} for LFQCD. An alternative  method to arrive at LF expressions would be to integrate over the light-front energy $k^-$ in covariant expressions. Paston {\it et al.} \citep{paston2} have considered the issue of equivalence between covariant QCD and light-front QCD at the Green's function level and have shown that equivalence between the two theories can be achieved  by adding non-conventional counterterms to the  LF Hamiltonian. The authors of Ref.\citep{paston2} mention that the LF Hamiltonian perturbation theory can be obtained from the LF Lagrangian perturbation theory by first integrating over $k^{-}$ and then over other components. In the present work, we consider the issue of equivalence at the Feynman diagram level in QED and show that the light-front expressions of the one loop diagrams of LFQED, derived using LF Hamiltonian perturbation theory \citep{mustaki, paston2}, can be obtained by performing $k^{-}$-integrations in the covariant expressions by carefully taking into account the contribution of end point singularities. 

\par Equivalence of covariant and LFQED, at one loop  level, was discussed in detail by one of us in Refs.\cite{swati,  swati2}, where it was shown that the one loop LF expressions
can be obtained by performing $k^-$-integration in the corresponding covariant expressions of equal-time theory.  Recently, the issue was revisited in  Ref.\cite{mantovani},  where the authors raised certain issues in our first work \citep{swati} and  also pointed out correctly  that our calculation for one loop vertex correction was only for the `$+$' component of $\Lambda^\mu$.

An important issue  in these  proofs of equivalence is that of the form of the gauge boson  propagator in light-front gauge, which has been a topic of keen interest in literature \cite{leibbrandt, leibbrandtrmp, suzukielsevier, suzukisalesarxiv04, jilisuzuki, mantovani, bhamremisra}. We consider this issue too in our present work. One loop renormalization of LFQED in LF gauge was discussed extensively in Ref.\cite{mustaki} using gauge boson propagator of the form \citep{kogutsoper} 
\begin{equation}
d_{\alpha\beta}(k)=-g_{\alpha\beta}+\frac{\delta_{\alpha+}k_{\beta}+\delta_{\beta+}k_{\alpha}}{k^+}
\end{equation}
which we shall refer to as the "two-term propagator" in this work. A method of arriving at this gauge propagator, without using the Cauchy Principal Value prescription to deal with the pole in the propagator, was developed using the gauge choice $A_{+}=0$ in Ref.\cite{soldatiqed} for QED and in Ref.\cite{soldatiym} for Yang-Mills theory. Srivastava and Brodsky, while discussing  the LF quantization of Hamiltonian QCD in detail, constructed S-matrix expansion in LF time-ordered products \citep{srivastavabrodsky}. These authors showed that the free field gauge boson propagator  is transverse with respect to both its 4-momentum and the gauge condition, and should have the form 
\begin{equation}
d'_{\alpha\beta}(k)=-g_{\alpha\beta}+\frac{\delta_{\alpha+}k_{\beta}+\delta_{\beta+}k_{\alpha}}{k^+}-\frac{\delta_{\alpha+}\delta_{\beta+}k^2}{(k^{+})^2}
\end{equation}
This form has been used in Refs.\citep{rohrlich, harindranath, yan, leibbrandt, leibbrandtrmp, suzukielsevier} and we  will refer to this doubly transverse gauge boson propagator as  the "three-term propagator"  in this work. Using a causal approach, it was shown in Ref.\cite{bufalofree} that the three-term propagator can be arrived at without making use of any specific prescription to handle the poles. Vacuum polarization calculation was performed using this approach in Ref.\cite{bufalovp}.

 Suzuki and Sales obtained, at the classical level, the gauge conditions that can lead to the three-term gauge boson propagator \cite{suzukielsevier, suzukisalesarxiv04}. The third term in this propagator is traditionally dropped on the grounds that it is exactly cancelled by the "instantaneous" term in the LF interaction Hamiltonian \citep{suzukisalesarxiv04} and it is argued that this term is unphysical and does  not propagate any information. However, the physical significance of this term has subsequently been stressed \cite{suzukielsevier}. It was shown in Ref.\citep{suzukielsevier} using the method of Lagrange's multiplier consistently that  the correct form of the gauge boson propagator necessarily has the third, contact term. The importance of this term in renormalization was also stressed by these authors. The equivalence of the manifestly covariant photon propagator to the sum of contributions from the transverse and longitudinal polarization of the virtual photon has been explicitly shown in Ref.\citep{jilisuzuki}.
\par

The issue of which form of photon propagator should be used in the proof of equivalence has been addressed by us \citep{swati, swati2} as well as by Mantovani {\it et al.} \citep{mantovani}. It was shown in Ref.\citep{mantovani} that the equivalence with the expressions of Mustaki {\it et al.} can be achieved using the method of performing $k^-$-integration only if one uses the two-term photon propagator. An important ingredient in this calculation consists of splitting the photon propagator into  on-shell and off-shell parts. In  Ref.\citep{swati2}, we had used an alternative method, called the Asymptotic Method proposed by Bakker {\it et al.} \citep{bakkerdewitt},  to prove equivalence and had shown  that, in case of vacuum polarization,  the instantaneous photon  exchange diagrams of  Ref.\citep{mustaki} can be generated  using the two-term propagator by carefully evaluating the contribution of the arc at infinity in contour integrations. 

\par

The present work  is motivated by the need to clarify the issue of form of photon propagator in LF gauge used in these proofs of equivalence. We present an alternative   proof of equivalence for fermion self-energy and vertex correction using the asymptotic method \cite{swati2} and also extend our earlier proof of equivalence to a general component of $\Lambda^\mu$. We stress the fact that the form of the propagator needed to achieve equivalence actually depends on whether one has used both the Lorentz condition and light-front gauge condition $A^+ = 0$ to arrive at the Light-front Hamiltonian or not.

\par  The starting point for our discussion is the equal-time QED Lagrangian in LF gauge given by 
\begin{equation} \label{eq:equaltime_QED_Lag}
{\cal L}=\frac{i}{2}\bar{\psi}\overset{\leftrightarrow}{\slashed{\partial}}\psi-m\bar{\psi}\psi-\frac{1}{4}F^{\mu\nu}F_{\mu\nu}-e\bar{\psi}\gamma^{\mu}\psi A_{\mu}-\frac{1}{2\alpha}(2n_{\mu}A^{\mu}) (\partial_{\nu}A^{\nu})
\end{equation}

\par

As shown by Suzuki {\it et al.}, this Lagrangian leads to the three-term propagator. The Lagrange's multiplier in the above Lagrangian takes care of the  Lorentz condition $ \partial_{\mu}A^{\mu} = 0 $ as well as the LF gauge condition $n_{\mu}A^{\mu} = 0 $. However, the Hamiltonian in Ref.\citep{mustaki} has been obtained by using only the LF gauge condition. Hence, in order to establish equivalence of covariant expressions with the results of Mustaki {\it et al.} \citep{mustaki}, it is appropriate only  to use  the two-term propagator in the covariant expression also, as  was done  by Mantovani {\it et al.} in Ref.\citep{mantovani},  since the third term of photon propagator arises from the Lorentz condition in ${\cal L}$. 

\par

 On the other hand, if we derive the LF Hamiltonian on the lines of Mustaki {\it et al.} but making use of the Lorentz condition as well  then,  as we will show in Sec. \ref {sec:propagator}, the 4-point vertex involving instantaneous photon exchange is not present in the Hamiltonian and hence the diagrams involving the instantaneous photon exchange will be absent in the one loop calculations. We establish equivalence of this theory with the covariant theory in subsequent sections.

\par

In Sec. \ref{sec:review}, we consider the fermion self-energy correction. We start with its covariant expression with the three-term propagator and show, by performing the $k^-$-integration, that indeed only the regular diagram and the instantaneous fermion exchange diagram of LFQED are generated using the methods of Mantovani {\it et al.}  For the sake of completeness, we also show that in the case of vacuum polarization too, the procedure of $k^-$-integration leads to the regular and instantaneous fermion exchange diagrams. 

\par

In Sec. \ref{sec:asymptotic}, we revisit, for the case of fermion self-energy, our earlier proof of equivalence using the Asymptotic Method \citep{bakkerdewitt, swati2}. We establish, using this method also, that the covariant expression with the two term-propagator, on performing the $k^-$-integration, leads to all the diagrams in Ref.\citep{mustaki}, while only regular and instantaneous fermion exchange diagrams are generated if the three-term propagator is used.

\par

In Sec. \ref{sec:equi_vertcorr}, we calculate the vertex correction contributions of the instantaneous fermion exchange diagrams in light-front QED that were not considered in our previous work \citep{swati} because of their matrix structure. This was  briefly discussed by us recently in Ref.\citep{bhamremisra}.

\par 
In Sec. \ref{sec:equi_vertcorrCov}, we establish equivalence between the covariant and LF expressions for a general component of  the one loop vertex correction $\Lambda^\mu$, by performing $k^{-}$-integration in the covariant expression. 

Finally, in Sec. \ref{sec:concl}, we summarize our results and comment on the issue of form of the photon proapgator. Appendix \ref{app:basics} contains the conventions and some basic formulae and Appendix \ref{app:details_sec_III} contains details of the calculations presented in Sec. \ref{sec:equi_vertcorr}.

\par

\section{Form of the photon propagator and the Light-Front Hamiltonian}\label{sec:propagator} 

There has been a great deal of discussion on the form of the photon propagator to be used in LF gauge as mentioned in the  Introduction.  Suzuki {\it et al.} have shown that classically the propagator derived from LF gauge Lagrangian in Eq.(\ref{eq:equaltime_QED_Lag}) has the third term also. Brodsky and Srivastava obtained this form in LF field theory and also showed that one necessarily has to introduce an instantaneous interaction term in the Hamiltonian if one eliminates the unphysical degrees of freedom.  We give below the form of the interaction Hamiltonian in this case for the sake of completeness \citep{mantovani}:
\begin{equation} \label{eq:L_lf_int}
{\cal H}_{int}=-{\cal L}_{LFint}=e\bar{\psi}\gamma^{\mu}\psi A_{\mu}
-\frac{e^{2}}{2}\bigg(\frac{1}{i\partial_{-}}\bar{\psi}\gamma^{+}\psi\bigg)\bigg(\frac{1}{i\partial_{-}}\bar{\psi}\gamma^{+}\psi\bigg)
\end{equation}
which is the QED analog of the LF QCD  Hamiltonian derived by Brodsky {\it et al.} \citep{srivastavabrodsky}. These authors have also shown that when the free gauge field satisfies both the Lorentz condition as well as the light-cone gauge condition, then its propagator is doubly transverse i.e. transverse to both its four-momentum and the gauge direction. 
The LF quantized QED Lagrangian in LF gauge in Eq.(\ref{eq:L_lf_int}) differs from the covariant form due to the presence of the second term representing an additional instantaneous interaction \citep{srivastavabrodsky}. As pointed out by Mantovani {\it et al.}, if one starts with the ${\cal L}_{LFint}$ in Eq.(\ref{eq:L_lf_int}), then the contribution of the  third term in the propagator cancels the contribution of the instantaneous vertex and therefore, it is sufficient to work with the two-term propagator. Thus the proof of equivalence as presented by Mantovani {\it et al.} deals with proving equivalence between Lagrangian formulation of LF quantized QED based on the Lagrangian in Eq.(\ref{eq:L_lf_int}) and the corresponding Hamiltonian version as given by Mustaki {\it et al.} \cite{mustaki}. 

\par

In this work, we investigate the issue of equivalence of equal-time quantized QED in standard covariant formulation based on the Lagrangian in Eq.(\ref{eq:equaltime_QED_Lag}) and LF quantized Hamiltonian QED as given in Eq.(\ref{eq:LF_Ham}) below. The main point that we stress in this work is  that the Lagrangian in Eq.(\ref{eq:equaltime_QED_Lag}) leads to a doubly transverse three-term propagator obtained using the fact that  the gauge field satisfies the Lorentz condition as well as the light-cone gauge condition, whereas the derivation of LF Hamiltonian in Ref.\citep{mustaki} uses only the LF gauge condition and the Lorentz condition is not taken into account. In the following, we re-visit their derivation, but taking into account the Lorentz condition as well, and show that the resulting Hamiltonian does not have the instantaneous photon interaction. 

We start with the QED Lagrangian 
\begin{equation}
{\cal L}=\frac{i}{2}\bar{\psi}\overset{\leftrightarrow}{\slashed\partial}\psi-m\bar{\psi}\psi-\frac{1}{4}F^{\mu\nu}F_{\mu\nu}-e\bar{\psi}\gamma^{\mu}\psi A_{\mu}
\end{equation}
which, after applying the light-cone gauge condition $A^+=0$, leads to the Light-Front Hamiltonian \citep{mustaki}
\[ P^- = P^{-}_G + P^{-}_F \]
where $P^{-}_G $ and  $P^{-}_F$ are bosonic and fermionic parts given by 
\begin{equation*}
P^{-}_G=\int d^{2}{\bf{x}}_{\perp}dx^{-}[(\partial_{-}A_{k})(\partial_{k}A_{+})-\frac{1}{2}(\partial_{-}A_{+})^{2}+\frac{1}{2}(F_{12})^{2}]
\end{equation*}
and
\begin{equation*}
P^{-}_F=\int d^{2}{\bf{x}}_{\perp}dx^{-}\bigg[\bar{\psi}\bigg[-\frac{i}{2}\gamma^{-}\overset{\leftrightarrow}\partial_{-}-\frac{i}{2}\gamma^{k}\overset{\leftrightarrow}\partial_{k}+m\bigg]\psi+J^{\mu}A_{\mu}\bigg]
\end{equation*}
(k=1,2)\\
Mustaki {\it et al.} obtained the Euler-Lagrange equation for $A_+$ which turns out be  a constraint equation
 \begin{equation} \label{eq:constraint_eq}
\partial^{2}_{-}A_{+}=\partial_{-}\partial_{k}A_{k}-J^{+}
\end{equation} 
using which  $A_+$ is eliminated. The Hamiltonian can then be expressed in terms of only physical transverse degrees of freedom of the photon and a non-local effective four-point vertex corresponding to instantaneous photon exchange is generated. 
 
\par

However, if one applies the light-cone gauge condition $ A^+=0$ as well as the Lorentz condition $ \partial \cdot A = 0 $, $A_+ $ does not appear in $P^{-}_G$  and  Eq.(\ref{eq:constraint_eq}) leads to $ J^+ = 0$. As a result, when the LF Hamiltonian is expressed in terms of only independent  degrees of  freedom, one obtains 
 \begin{equation} \label{eq:LF_Ham}
P^{-}=H_{0}+V_{1}+V_{2}
\end{equation}
Here,
\begin{equation*}
H_{0}=\int d^{2}{\bf{x}}_{\perp}dx^{-}\big[\frac{i}{2}\bar{\xi}\gamma^{-}\overset{\leftrightarrow}\partial_{-}\xi+\frac{1}{2}(F_{12})^{2}-\frac{1}{2}a_{+}\partial_{-}\partial_{k}a_{k}\big]
\end{equation*}
is the free Hamiltonian, 
\begin{equation*} \label{eq:V1_expression}
V_1=e\int d^{2}{\bf{x}}_{\perp}dx^{-}\bar\xi\gamma^{\mu}\xi a_{\mu}
\end{equation*}
is the standard, order-$e$, three-point interaction, and 
\begin{equation*} \label{eq:V2_expression}
V_{2}=-\frac{i}{4}e^{2}\int d^{2}{\bf{x}}_{\perp}dx^{-}dy^{-}\epsilon(x^{-}-y^{-})(\bar{\xi}a_{k}\gamma^{k})(x)\gamma^{+}(a_{j}\gamma^{j}\xi)(y)
\end{equation*}
is order-$e^2$ non-local four-point vertex corresponding to an instantaneous fermion exchange.

\par
  
Note that this Hamiltonian differs from the Hamiltonian obtained by Mustaki {\it et al.} by the absence of the  non-local instantaneous photon exchange interaction.  The non-local interaction involving instantaneous fermion exchange is still present though. In the next section, we will draw all the basic one loop graphs in  LFQED resulting from this Hamiltonian. The expressions for these diagrams were obtained in Ref.\cite{mustaki}. We will then show that these can all be generated by performing $k^-$-integration in the covariant expressions with the three-term photon propagator.

  \section{Equivalence of Covariant and Light-Front one loop expressions}\label{sec:review} 
  
  In this section, we first give expressions for one loop corrections in our formulation of LFQED which have been obtained using the techniques of old fashioned time-ordered perturbation theory in the light-front framework. The one loop diagrams considered here are a subset of diagrams given in Ref.\cite{mustaki} due to the absence of instantaneous photon exchange interaction  in our formulation. We will then show that these expressions can be obtained from the corresponding covariant expressions by performing $k^-$-integration consistently while using the three-term photon propagator.  We use the method of performing the $k^-$-integrations \citep{swati} to establish equivalence. In this section, we use the procedure followed in Ref.\citep{mantovani} for dealing with the divergences coming from the photon propagator. The results will be reproduced in Sec. \ref{sec:asymptotic} using the asymptotic method.  In Sec. \ref{sec:equi_self-en}, we show the equivalence of one loop fermion self-energy graph in covariant QED described by the Lagrangian in Eq.(\ref{eq:equaltime_QED_Lag}) with the light-front QED fermion self-energy graphs resulting from the Hamiltonian in Eq.(\ref{eq:LF_Ham}). As explained in Sec. \ref{sec:propagator}, we use the three-term propagator in place of the two-term propagator and compare the results with those in Ref.\citep{mantovani}. Sec. \ref{sec:equi_vacpol} is a review of the work done in Ref.\citep{swati}. This is included for the sake of completeness. We defer the proof of equivalence for vertex correction to Sec. \ref{sec:equi_vertcorrCov} till after calculating the one loop vertex correction in Sec. \ref{sec:equi_vertcorr} for a general component of $\Lambda^\mu$.  

  \subsection{Fermion Self-Energy}\label{sec:equi_self-en}
  
\begin{figure}[h!]
\centering
\subfloat[]
{\includegraphics[scale=0.65]{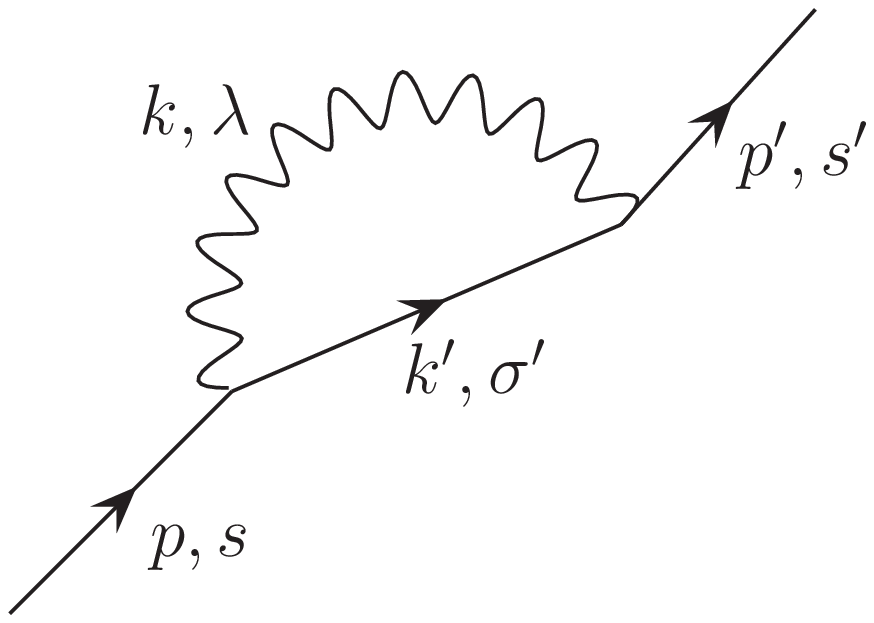}}
\hspace*{1cm}
\subfloat[]
{\includegraphics[scale=0.65]{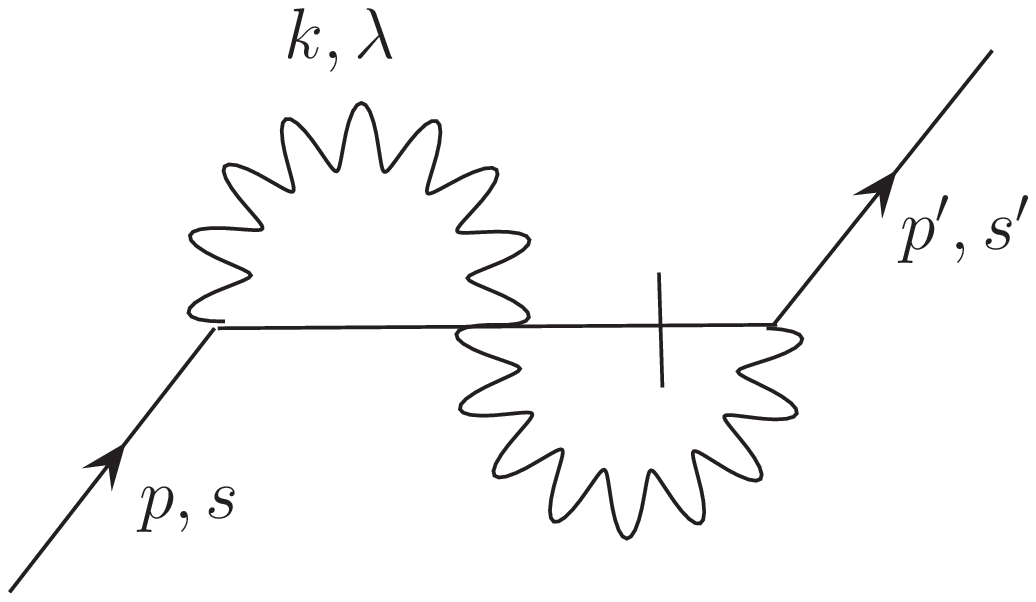}}
\caption{"Regular" and instantaneous fermion exchange self-energy diagrams}
\label{fig:selfen}
\end{figure}

 Starting with the LF Hamiltonian in Eq.(\ref{eq:LF_Ham}),  one obtains the one loop corrections to the fermion self-energy in the light-front time-ordered perturbation theory which consist of the "regular" diagram and an instantaneous fermion exchange diagram  shown in Fig.(\ref{fig:selfen}).

Following the procedure in Ref.\citep{mustaki}, one obtains the following expressions for these diagrams which are Eqs.(3.9)-(3.10) in Ref.\cite{mustaki}:
\begin{equation} \label{eq:se_reg}
\bar{u}_{p',s'}\Sigma_{1}u_{p,s}=\frac{e^{2}}{m}\int\frac{d^{2}\bf{k}_{\perp}}{(4\pi)^{3}}\int_{0}^{p^{+}}\frac{dk^{+}}{k^{+}(p^{+}-k^{+})}\frac{\bar{u}_{p,s'}\gamma^{\mu}({k\llap/}'+m)\gamma^{\nu}u_{p,s}d_{\mu\nu}(k)}{p^{-}-k^{-}-k'^{-}}
\end{equation}
for the regular diagram, and 
\begin{equation} \label{eq:se_inst_ferm1}
\bar{u}_{p',s'}\Sigma_{2}u_{p,s}=\frac{e^{2}p^{+}\delta_{ss'}}{2m}\int\frac{d^{2}\bf{k}_{\perp}}{(2\pi)^{3}}\int_{0}^{\infty}\frac{dk^{+}}{k^{+}(p^{+}-k^{+})}
\end{equation}
for the instantaneous fermion exchange diagram.

In the standard covariant formulation of equal-time quantized QED, only Fig.(\ref{fig:selfen})(a) is present and the expression in light-front gauge is
\begin{equation} \label{eq:se_main1}
\Sigma(p)=\frac{ie^{2}}{2m}\int\frac{d^{4}k}{(2\pi)^{4}}\frac{{\gamma^\mu}{({p\llap/-k\llap/}+m)}{\gamma^\nu}d'_{\mu\nu}{(k)}}{[(p-k)^2-m^2+i\epsilon][k^2-{\mu}^2+i\epsilon]}
\end{equation}
where
\begin{equation*}
d^{\prime}_{\alpha\beta}(k)=d_{\alpha\beta}(k)-\frac{\delta_{\alpha+}\delta_{\beta+}k^2}{(k^{+})^2}=-g_{\alpha\beta}+\frac{\delta_{\alpha+}k_{\beta}+\delta_{\beta+}k_{\alpha}}{k^+}-\frac{\delta_{\alpha+}\delta_{\beta+}k^2}{(k^{+})^2}
\end{equation*}
To show equivalence of covariant and LF expressions, one rewrites $p\llap/-k\llap/$ in Eq.(\ref{eq:se_main1}) as a sum of an on-shell part and an off-shell part \citep{swati}:
\begin{equation}\label{eq:splitting_1}
\begin{split}
p\llap/-k\llap/ =& \gamma^{+}(p^{-}-k^{-})+\gamma^{-}(p^{+}-k^{+})+\gamma^{\perp}({\bf{p}}_{\perp}-{\bf{k}}_{\perp})\\
=& \gamma^{+}\bigg[\frac{({\bf{p}}_{\perp}-{\bf{k}}_{\perp})^{2}+m^{2}}{2(p^{+}-k^{+})}\bigg]+\gamma^{-}(p^{+}-k^{+})+\gamma^{\perp}({\bf{p}}_{\perp}-{\bf{k}}_{\perp})+\gamma^{+}\bigg[(p^{-}-k^{-})-\bigg[\frac{({\bf{p}}_{\perp}-{\bf{k}}_{\perp})^{2}+m^{2}}{2(p^{+}-k^{+})}\bigg]\bigg]\\
=& k\llap/'_{on}+\frac{\gamma^{+}[(p-k)^{2}-m^{2}]}{2(p^{+}-k^{+})}
\end{split}
\end{equation}
which leads to
\begin{equation*}
\Sigma(p)=\Sigma_{1}(p)+\Sigma_{2}(p)
\end{equation*}
where
\begin{equation*}
\begin{split}
\Sigma_{1}(p)=& \frac{ie^{2}}{2m}\int\frac{d^{4}k}{(2\pi)^{4}}\frac{\gamma^{\mu}(k\llap/'_{on}+m)\gamma^{\nu}d_{\mu\nu}(k)}{[(p-k)^{2}-m^{2}+i\epsilon][k^{2}-\mu^{2}+i\epsilon]}\\
-&\frac{ie^{2}}{2m}\int\frac{d^{4}k}{(2\pi)^{4}}\frac{\gamma^{\mu}(k\llap/'_{on}+m)\gamma^{\nu}\delta_{\mu+}\delta_{\nu+}k^{2}}{[(p-k)^{2}-m^{2}+i\epsilon][k^{2}-\mu^{2}+i\epsilon](k^{+})^{2}}
\end{split}
\end{equation*}
and
\begin{equation*}
\Sigma_{2}(p)=\frac{ie^{2}}{2m}\int\frac{d^{4}k}{(2\pi)^{4}}\frac{\gamma^{\mu}\gamma^{+}\gamma^{\nu}d'_{\mu\nu}(k)}{2(p^{+}-k^{+})[k^{2}-\mu^{2}+i\epsilon]}
\end{equation*}
Using the identities $\gamma^{\alpha}\gamma^{\mu}\gamma^{\beta}d_{\alpha\beta}(k)=\frac{2}{k^{+}}[\gamma^{+}k^{\mu}+g^{+\mu}k\llap/]$, $(\gamma^{+})^{2}=0$ and $\bar{u}_{p,s'}\gamma^{\mu}u_{p,s}=2p^{\mu}\delta_{ss'}$, the last expression leads to
\begin{equation*}
\bar{u}_{p,s'}\Sigma_{2}(p)u_{p,s}=\frac{ie^{2}p^{+}\delta_{ss'}}{2m}\int\frac{d^{2}{\bf{k}}_{\perp}dk^{+}}{(2\pi)^{4}k^{+}(p^{+}-k^{+})}\int\frac{dk^{-}}{\big(k^{-}-\frac{{\bf{k}}^{2}_{\perp}+\mu^{2}-i\epsilon}{2k^{+}}\big)}
\end{equation*}
The $k^{-}$-integral in this equation has a pole at $k^{-}_{1}=\frac{{\bf{k}}^{2}_{\perp}+\mu^{2}-i\epsilon}{2k^{+}}$ which approaches infinity as $k^+ \rightarrow 0$. In order to deal with the pole at infinity, we use the method of $u$-integration \citep{bakkerdewitt,swati}. We make the change of variable $u=\frac{1}{k^{-}}$ thus modifying the integral to
\begin{equation}
\int_{-\infty}^{+\infty}\frac{du}{u\big[1-u\big(\frac{{\bf{k}}^{2}_{\perp}+\mu^{2}-i\epsilon}{2k^{+}}\big)\big]}
\end{equation}
The $u$-integral needs to be regulated and hence we write $\frac{1}{u}=\frac{1}{2}\big[\frac{1}{u+i\delta}+\frac{1}{u-i\delta}\big]$ which leads to
\begin{equation*}
\int\frac{dk^{-}}{\big(k^{-}-\frac{{\bf{k}}^{2}_{\perp}+\mu^{2}-i\epsilon}{2k^{+}}\big)}=\frac{1}{2}\int_{-\infty}^{+\infty}\frac{du}{(u+i\delta)\big[1-u\big(\frac{{\bf{k}}^{2}_{\perp}+\mu^{2}-i\epsilon}{2k^{+}}\big)\big]}+\frac{1}{2}\int_{-\infty}^{+\infty}\frac{du}{(u-i\delta)\big[1-u\big(\frac{{\bf{k}}^{2}_{\perp}+\mu^{2}-i\epsilon}{2k^{+}}\big)\big]}
\end{equation*}
In the above equation, the first $u$-integral has poles at $u_{1}=-i\delta$ and $u_{2}=\frac{2k^{+}}{{\bf{k}}^{2}_{\perp}+\mu^{2}-i\epsilon}$. For $k^{+}<0$, the integral is zero since both poles lie below the real axis. For $k^{+}>0$, $u_{1}$ lies below and $u_{2}$ above the real axis. Closing the contour below gives the value of the integral as $-2\pi i\theta(k^{+})$ as $\delta\rightarrow0$. The $u$-integral in the second term has poles at $u_{1}=i\delta$ and $u_{2}=\frac{2k^{+}}{{\bf{k}}^{2}_{\perp}+\mu^{2}-i\epsilon}$. For $k^{+}>0$, both poles lie above the real axis and the integral vanishes on closing the contour in the lower half-plane. For $k^{+}<0$, we close the contour in the upper half-plane as $u_{1}$ lies above and $u_{2}$ below the real line. The value of the integral is $2\pi i\theta(-k^{+})$ as $\delta\rightarrow0$. Thus,
\begin{equation*}
\int\frac{dk^{-}}{\big(k^{-}-\frac{{\bf{k}}^{2}_{\perp}+\mu^{2}-i\epsilon}{2k^{+}}\big)}=-\pi i\theta(k^{+})+\pi i\theta(-k^{+})
\end{equation*}
and
\begin{equation*}
\bar{u}_{p,s'}\Sigma_{2}(p)u_{p,s}=\frac{e^{2}p^{+}\delta_{ss'}}{2m}\int\frac{d^{2}{\bf{k}}_{\perp}}{(2\pi)^{3}}\bigg[\frac{1}{2}\int_{0}^{\infty}\frac{dk^{+}}{k^{+}(p^{+}-k^{+})}-\frac{1}{2}\int_{-\infty}^{0}\frac{dk^{+}}{k^{+}(p^{+}-k^{+})}\bigg]
\end{equation*}
which can be shown to be the  same as
\begin{equation} \label{eq:se_sigma2}
\bar{u}_{p,s'}\Sigma_{2}(p)u_{p,s}=\frac{e^{2}p^{+}\delta_{ss'}}{2m}\int\frac{d^{2}{\bf{k}}_{\perp}}{(2\pi)^{3}}\int_{0}^{\infty}\frac{dk^{+}}{k^{+}(p^{+}-k^{+})}
\end{equation}
This is nothing but the expression for instantaneous fermion exchange diagram  as given in Eq.(\ref{eq:se_inst_ferm1}). 

$\Sigma_{1}(p)$ can be written as 
\[ \Sigma_{1}(p) = \Sigma_{1}^{(a)}(p) +\Sigma_{1}^{(b)}(p) \]
where 
\begin{equation} \label{eq:sigma_1-a}
\Sigma_{1}^{(a)}(p)= \frac{ie^{2}}{2m}\int\frac{d^{4}k}{(2\pi)^{4}}\frac{\gamma^{\mu}(k\llap/'_{on}+m)\gamma^{\nu}d_{\mu\nu}(k)}{[(p-k)^{2}-m^{2}+i\epsilon][k^{2}-\mu^{2}+i\epsilon]}
\end{equation}
and 
\begin{equation}\label{eq:sigma_1-b}
\Sigma_{1}^{(b)}(p)=- \frac{ie^{2}}{2m}\int\frac{d^{4}k}{(2\pi)^{4}}\frac{\gamma^{\mu}(k\llap/'_{on}+m)\gamma^{\nu}\delta_{\mu+}\delta_{\nu+}k^{2}}{[(p-k)^{2}-m^{2}+i\epsilon][k^{2}-\mu^{2}+i\epsilon](k^{+})^{2}}
\end{equation}
Using the identities $\gamma^{+}\gamma^{-}\gamma^{+}=2\gamma^{+}$, $(\gamma^{+})^{2}=0$ and $\bar{u}_{p,s'}\gamma^{\mu}u_{p,s}=2p^{\mu}\delta_{ss'}$, we get
\begin{equation*} \label{eq:se_sigma1b}
\bar{u}_{ps'}\Sigma_{1}^{(b)}(p)u_{ps}= -\frac{2ie^{2}p^{+}\delta_{ss'}}{2m}\int\frac{d^{2}{\bf{k}}_{\perp}dk^{+}}{(2\pi)^{4}(k^{+})^{2}}\int\frac{dk^{-}}{\big[p^{-}-k^{-}-\frac{(p_{\perp}-{\bf{k}}_{\perp})^{2}+m^{2}-i\epsilon}{2(p^{+}-k^{+})}\big]}
\end{equation*}
The $k^{-}$-integral in this  equation has a pole at $k^{-}_{1}=p^{-}-\frac{(p_{\perp}-{\bf{k}}_{\perp})^{2}+m^{2}-i\epsilon}{2(p^{+}-k^{+})}$ and at infinity as $k^+\rightarrow p^{+}$. Evaluating the $k^{-}$-integral along the same lines as in case of $\Sigma_{2}(p)$, we get
\begin{equation*}
\int\frac{dk^{-}}{\big(p^{-}-k^{-}-\frac{(p_\perp-{\bf{k}}_\perp)^{2}+m^{2}-i\epsilon}{2(p^{+}-k^{+})}\big)}=-\pi i\theta(p^{+}-k^{+})+\pi i\theta(k^{+}-p^{+})
\end{equation*}
Thus,
\begin{equation}
\bar{u}_{ps'}\Sigma_{1}^{(b)}(p)u_{ps}= \frac{e^{2}p^{+}\delta_{ss'}}{2m}\int\frac{d^{2}{\bf{k}}_{\perp}}{(2\pi)^{3}}\bigg[\int_{p^{+}}^{\infty}\frac{dk^{+}}{(k^{+})^{2}}-\int_{-\infty}^{p^{+}}\frac{dk^{+}}{(k^{+})^{2}}\bigg]
\end{equation}
Changing the variable $k^{+}\rightarrow(p^{+}+k^{+})$ in the first integral and $k^{+}\rightarrow(p^{+}-k^{+})$ in the second gives
\begin{equation}\label{eq:sigma_1b}
\bar{u}_{p,s'}\Sigma_{1}^{(b)}(p)u_{p,s} = \frac{e^{2}p^{+}\delta_{ss'}}{2m}\int\frac{d^{2}{\bf{k}}_{\perp}}{(2\pi)^{3}}\bigg[\int_{0}^{\infty}\frac{dk^{+}}{(p^{+}+k^{+})^{2}}-\int_{0}^{\infty}\frac{dk^{+}}{(p^{+}-k^{+})^{2}}\bigg]
\end{equation}

$\Sigma_{1}^{(a)}(p)$ can be evaluated using the method of splitting $d_{\mu \nu}$ into  on-shell and off-shell parts as done in Ref.\citep{mantovani}. This leads, after  performing the $k^{-}$-integration, to the following two expressions:
\begin{equation}\label{eq:sigma_1a1}
\bar{u}_{p,s'}\Sigma_{1}^{(a1)}(p)u_{p,s} = \frac{e^{2}}{m}\int\frac{d^{2}{\bf{k}}_{\perp}}{(4\pi)^{3}}\int_{0}^{p^{+}}\frac{dk^{+}}{k^{+}(p^{+}-k^{+})}\frac{\bar{u}_{p,s'}\gamma^{\mu}(k\llap/'_{on}+m)\gamma^{\nu}u_{p,s}d_{\mu\nu}(k_{on})}{p^{-}-k^{-}_{on}-k'^{-}_{on}}
\end{equation}
and
\begin{equation}\label{eq:sigma_1a2}
\bar{u}_{p,s'}\Sigma_{1}^{(a2)}(p)u_{p,s} = -\frac{e^{2}p^{+}\delta_{ss'}}{2m}\int\frac{d^{2}{\bf{k}}_{\perp}}{(2\pi)^{3}}\bigg[\int_{0}^{\infty}\frac{dk^{+}}{(p^{+}+k^{+})^{2}}-\int_{0}^{\infty}\frac{dk^{+}}{(p^{+}-k^{+})^{2}}\bigg]
\end{equation}
Eq.(\ref{eq:sigma_1a1}) above, which is the same as Eq.(57) in Ref.\citep{mantovani}, is the expression for the "regular" diagram and  Eq.(\ref{eq:sigma_1a2}) cancels the contribution of Eq.(\ref{eq:sigma_1b}).

\par

\begin{figure}[h!]
\centering
\includegraphics[scale=0.65]{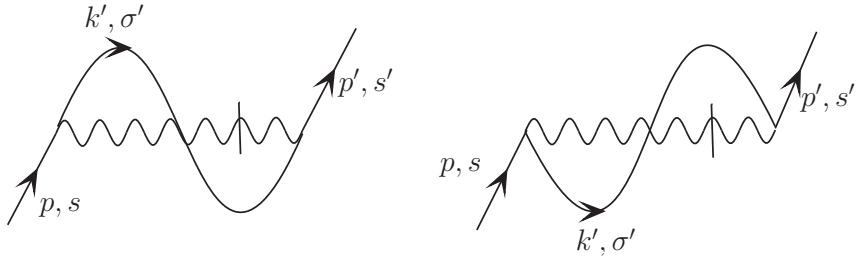}
\caption{Instantaneous photon exchange self-energy diagram}
\label{fig:selfen_inst_photon}
\end{figure}

In conclusion, using the three-term photon propagator and employing the method of splitting the propagator into on-shell and off-shell parts as given in Ref.\citep{mantovani}, the "regular" and instantaneous fermion exchange diagrams are generated by performing $k^{-}$-integration in the covariant expression. Had we started with the two-term propagator instead, as was done by Mantovani {\it et al.}, Eq.(\ref{eq:sigma_1-b}) and hence Eq.(\ref{eq:sigma_1b}) would have been  absent leaving Eq.(\ref{eq:sigma_1a2}) intact, which, in fact,  is the expression for the instantaneous photon diagram given in Fig.(\ref{fig:selfen_inst_photon}). This diagram is not present in our formulation based on the Hamiltonian in  Eq.(\ref{eq:LF_Ham}). These findings are consistent with the discussion in Sec. \ref{sec:propagator}.

\par In the next section, we will employ an alternative method called the Asymptotic Method, in place  of splitting the propagator into on-shell and off-shell parts, to achieve the same results. 

\subsection{Vacuum Polarization}\label{sec:equi_vacpol}

\begin{figure}[h!]
\centering
\subfloat[]
{\includegraphics[scale=0.5]{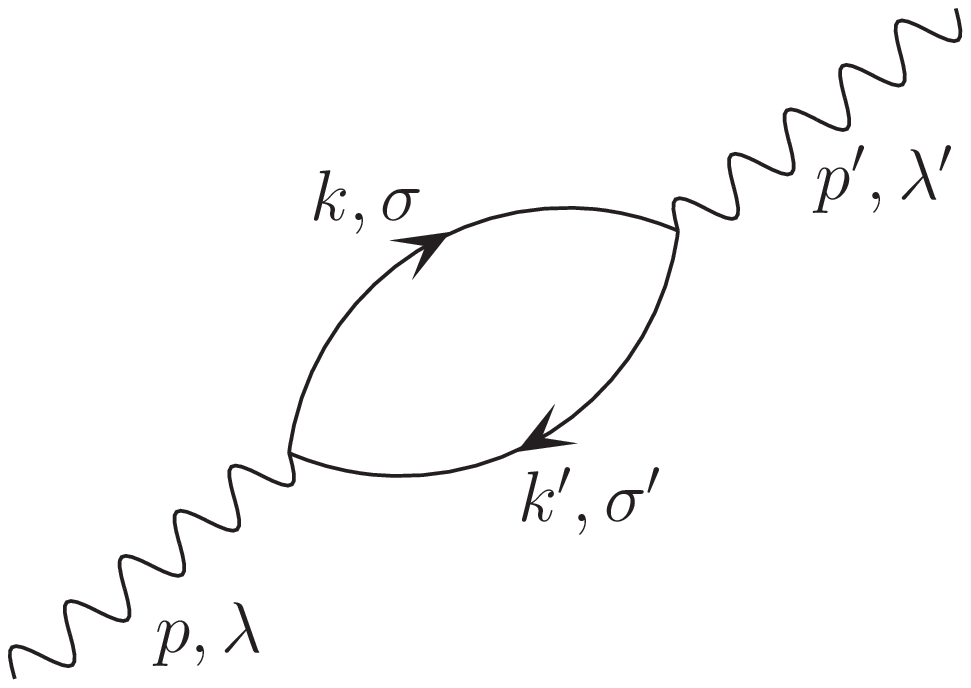}}
\hspace{0.7cm}
\subfloat[]
{\includegraphics[scale=0.5]{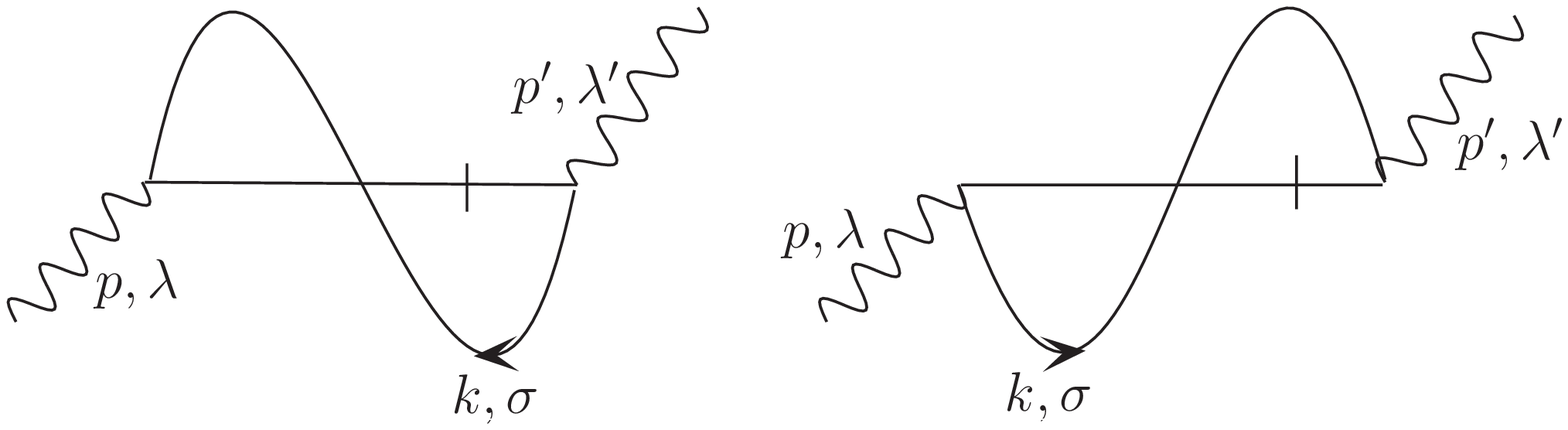}}
\caption{"Regular" and instantaneous vacuum polarization diagrams}
\label{fig:vacpol}
\end{figure}

The  one loop diagrams that  contribute to vacuum polarization in LFTOPT are given in Fig.(\ref{fig:vacpol}). The expressions for these diagrams, as given in Eqs.(4.6) and (4.5) of Ref.\cite{mustaki}, are as follows:
\begin{equation} \label{eq:vp_reg1}
\epsilon^{\lambda}_{\mu}(p)\Pi^{\mu\nu}_{1}\epsilon^{\lambda'}_{\nu}(p)=2e^{2}\int\frac{d^{2}{\bf{k}}_{\perp}}{(4\pi)^{3}}\int_{0}^{p^{+}}\frac{dk^{+}}{k^{+}(p^{+}-k^{+})}\frac{Tr[\epsilon\llap/^{\lambda}(p)(k\llap/+m)\epsilon\llap/^{\lambda'}(p)(k\llap/'-m)]}{p^{-}-k^{-}-k'^{-}}
\end{equation}
for the "regular" diagram, and
\begin{equation} \label{eq:vp_inst1}
\epsilon^{\lambda}_{\mu}(p)\Pi^{\mu\nu}_{2}\epsilon^{\lambda}_{\nu}(p)=e^{2}\int\frac{d^{2}{\bf{k}}_{\perp}}{(2\pi)^{3}}\int_{0}^{\infty}dk^{+}\bigg[\frac{1}{p^{+}-k^{+}}-\frac{1}{p^{+}+k^{+}}\bigg]
\end{equation}
for the two instantaneous diagrams.

The standard covariant expression for vacuum polarization is given by
\begin{equation} \label{eq:vp_main1}
\Pi^{\mu\nu}(p)=ie^{2}\int\frac{d^{4}k}{(2\pi)^{4}}\frac{Tr[\gamma^{\mu}(k\llap/+m)\gamma^{\nu}(p\llap/-k\llap/-m)]}{(k^{2}-m^{2}+i\epsilon)[(p-k)^{2}-m^{2}+i\epsilon]}
\end{equation}
Employing a similar scheme as in the case of self-energy, we split $(k\llap/+m)$ and $(p\llap/-k\llap/-m)$ in the above equation into on-shell and off-shell parts i.e. $k\llap/+m=k\llap/_{on}+m+\frac{\gamma^{+}(k^{2}-m^{2})}{2k^{+}}$, $p\llap/-k\llap/-m=k\llap/'_{on}-m+\frac{\gamma^{+}[(p-k)^{2}-m^{2}]}{2(p^{+}-k^{+})}$. Doing so, Eq.(\ref{eq:vp_main1}) splits up as follows:
\begin{equation}
\Pi^{\mu\nu}(p)=\Pi^{\mu\nu}_{1}(p)+\Pi^{\mu\nu}_{2}(p)+\Pi^{\mu\nu}_{3}(p)+\Pi^{\mu\nu}_{4}(p)
\end{equation}
where
\begin{equation*}
\Pi^{\mu\nu}_{1}(p)=ie^{2}\int\frac{d^{4}k}{(2\pi)^{4}}\frac{Tr[\gamma^{\mu}(k\llap/_{on}+m)\gamma^{\nu}(k\llap/'_{on}-m)]}{2k^{+}\ 2(p^{+}-k^{+})\big(k^{-}-\frac{{\bf{k}}^{2}_{\perp}+m^{2}-i\epsilon}{2k^{+}}\big)\big(p^{-}-k^{-}-\frac{({\bf{p}}_{\perp}-{\bf{k}}_{\perp})^{2}+m^{2}-i\epsilon}{2(p^{+}-k^{+})}\big)},
\end{equation*}

\begin{equation*}
\Pi^{\mu\nu}_{2}(p)=ie^{2}\int\frac{d^{4}k}{(2\pi)^{4}}\frac{Tr[\gamma^{\mu}(k\llap/_{on}+m)\gamma^{\nu}\gamma^{+}]}{2k^{+}\ 2(p^{+}-k^{+})\big(k^{-}-\frac{{\bf{k}}^{2}_{\perp}+m^{2}-i\epsilon}{2k^{+}}\big)},
\end{equation*}

\begin{equation*}
\Pi^{\mu\nu}_{3}(p)=ie^{2}\int\frac{d^{4}k}{(2\pi)^{4}}\frac{Tr[\gamma^{\mu}\gamma^{+}\gamma^{\nu}(k\llap/'_{on}-m)]}{2k^{+}\ 2(p^{+}-k^{+})\big(p^{-}-k^{-}-\frac{({\bf{p}}_{\perp}-{\bf{k}}_{\perp})^{2}+m^{2}-i\epsilon}{2(p^{+}-k^{+})}\big)},
\end{equation*}

\begin{equation*}
\Pi^{\mu\nu}_{4}(p)=ie^{2}\int\frac{d^{4}k}{(2\pi)^{4}}\frac{Tr[\gamma^{\mu}\gamma^{+}\gamma^{\nu}\gamma^{+}]}{2k^{+}\ 2(p^{+}-k^{+})}
\end{equation*}
On using the fact that $\epsilon_{-}=0$, the identity $(\gamma^{+})^{2}=0$, and the anticommutation relations of $\gamma$-matrices, we can see that the contribution of $\Pi^{\mu\nu}_{4}(p)$ to the transition amplitude viz. $\epsilon^{\lambda}_{\mu}\Pi^{\mu\nu}_{4}(p)\epsilon^{\lambda'}_{\nu}$ is null.

\par 

The $k^{-}$-integral of $\Pi^{\mu\nu}_{1}(p)$ i.e.
\begin{equation*}
\int\frac{dk^{-}}{\big(k^{-}-\frac{{\bf{k}}^{2}_{\perp}+m^{2}-i\epsilon}{2k^{+}}\big)\big(p^{-}-k^{-}-\frac{({\bf{p}}_{\perp}-{\bf{k}}_{\perp})^{2}+m^{2}-i\epsilon}{2(p^{+}-k^{+})}\big)}
\end{equation*}
has poles at $k^{-}_{1}=\frac{{\bf{k}}^{2}_{\perp}+m^{2}-i\epsilon}{2k^{+}}$ and at $k^{-}_{2}=p^{-}-\frac{({\bf{p}}_{\perp}-{\bf{k}}_{\perp})^{2}+m^{2}-i\epsilon}{2(p^{+}-k^{+})}$. For $k^{+}<0$, both poles lie above the real axis and for $k^{+}>p^{+}$, both lie below it. Hence, on closing the contour on the opposite side of the position of poles, both the ranges  $k^{+}<0$ and $k^{+}>p^{+}$ provide no contribution to the integral. For $0<k^{+}<p^{+}$, $k^{-}_{1}$ lies below the real axis and $k^{-}_{2}$ lies above. Closing the contour below, the value of this $k^{-}$-integral is
\begin{equation*}
\frac{-2\pi i\theta(k^{+})\theta(p^{+}-k^{+})}{\big(p^{-}-\frac{{\bf{k}}^{2}_{\perp}+m^{2}-i\epsilon}{2k^{+}}-\frac{({\bf{p}}_{\perp}-{\bf{k}}_{\perp})^{2}+m^{2}-i\epsilon}{2(p^{+}-k^{+})}\big)}
\end{equation*}
and hence
\begin{equation} \label{eq:vp_pi1}
\epsilon^{\lambda}_{\mu}(p)\Pi^{\mu\nu}_{1}\epsilon^{\lambda'}_{\nu}(p)=2e^{2}\int\frac{d^{2}{\bf{k}}_{\perp}}{(4\pi)^{3}}\int_{0}^{p^{+}}\frac{dk^{+}}{k^{+}(p^{+}-k^{+})}\frac{Tr[\epsilon\llap/^{\lambda}(p)(k\llap/_{on}+m)\epsilon\llap/^{\lambda'}(p)(k\llap/'_{on}-m)]}{p^{-}-k^{-}_{on}-k'^{-}_{on}}
\end{equation}
which is same as the expression for the diagram in Fig.(\ref{fig:vacpol})(a) given by Eq.(\ref{eq:vp_reg1}).

\par 



Using the trace properties, $\Pi^{\mu\nu}_{2}(p)$ further reduces to
\begin{equation*}
\Pi^{\mu\nu}_{2}(p)=ie^{2}\int\frac{d^{4}k}{(2\pi)^{4}}\frac{k^{-}_{on}Tr[\gamma^{\mu}\gamma^{+}\gamma^{\nu}\gamma^{+}]+k^{+}Tr[\gamma^{\mu}\gamma^{-}\gamma^{\nu}\gamma^{+}]}{2k^{+}\ 2(p^{+}-k^{+})\big(k^{-}-\frac{{\bf{k}}^{2}_{\perp}+m^{2}-i\epsilon}{2k^{+}}\big)}
\end{equation*}
The first term of the numerator in the above integral provides no contribution to $\epsilon^{\lambda}_{\mu}(p)\Pi^{\mu\nu}_{2}\epsilon^{\lambda}_{\nu}(p)$ since $(\gamma^{+})^{2}=0$, $\epsilon_{-}=0$ and $\{\gamma^{i},\gamma^{+}\}=0$. Thus,
\begin{equation*}
\epsilon^{\lambda}_{\mu}(p)\Pi^{\mu\nu}_{2}\epsilon^{\lambda}_{\nu}(p)=ie^{2}\int\frac{d^{2}{\bf{k}}_{\perp}dk^{+}}{(2\pi)^{4}}\frac{\epsilon^{\lambda}_{\mu}Tr[\gamma^{\mu}\gamma^{-}\gamma^{\nu}\gamma^{+}]\epsilon^{\lambda}_{\nu}}{4(p^{+}-k^{+})}\int\frac{dk^{-}}{\big(k^{-}-\frac{{\bf{k}}^{2}_{\perp}+m^{2}-i\epsilon}{2k^{+}}\big)}
\end{equation*}
The numerator in the above integral, $\epsilon^{\lambda}_{\mu}Tr[\gamma^{\mu}\gamma^{-}\gamma^{\nu}\gamma^{+}]\epsilon^{\lambda}_{\nu}=4$, which can be shown  using $\epsilon^{\lambda}_{\mu}p^{\mu}=0$, $\epsilon^{\lambda'}_{\mu}\epsilon^{\lambda\mu}=-\delta_{\lambda'\lambda}$, $\epsilon^{\lambda}_{-}=0$.
The $k^{-}$-integral is the same as  evaluated in the previous subsection. Thus, we have,
\begin{equation} \label{eq:vp_pi2}
\epsilon^{\lambda}_{\mu}(p)\Pi^{\mu\nu}_{2}\epsilon^{\lambda}_{\nu}(p)=\frac{e^{2}}{2}\int\frac{d^{2}{\bf{k}}_{\perp}}{(2\pi)^{3}}\int_{0}^{\infty}dk^{+}\bigg[\frac{1}{p^{+}-k^{+}}-\frac{1}{p^{+}+k^{+}}\bigg]
\end{equation}
The calculation of $\Pi^{\mu\nu}_{3}$ follows exactly on the lines of $\Pi^{\mu\nu}_{2}$ and we have
\begin{equation} \label{eq:vp_pi3}
\epsilon^{\lambda}_{\mu}(p)\Pi^{\mu\nu}_{3}\epsilon^{\lambda}_{\nu}(p)=\frac{e^{2}}{2}\int\frac{d^{2}{\bf{k}}_{\perp}}{(2\pi)^{3}}\int_{0}^{\infty}dk^{+}\bigg[\frac{1}{p^{+}-k^{+}}-\frac{1}{p^{+}+k^{+}}\bigg]
\end{equation}
The above two equations add up to give the light-front expression for instantaneous fermion diagrams i.e. Eq.(\ref{eq:vp_inst1}). It can be inferred from the above calculations that\\
(i) the regular diagram in LFTOPT corresponds to the situation where both the fermions in the loop are on-shell, and\\
(ii) the additional (instantaneous) diagrams that contribute to the photon self-energy can be looked at as being the result of one of the fermions going off-shell. 

\section{The Asymptotic Method}\label{sec:asymptotic}

Asymptotic method was introduced in Ref. {\citep{bakkerdewitt} by Bakker {\it et al.} in the context of $(1+1)-$ theories as an alternative to explicit evaluation of  arc contribution. In this method, one deals directly with the linear divergences as $k^+ \rightarrow 0$ and as $k^+ \rightarrow p^+$ and  isolates the divergent part  by evaluating the integrand at the asymptotic values of $k^-$. The method was used by us in Ref.\citep{swati2} in the context of QED.  In this section, we will employ the Asymptotic Method  first using  the two-term gauge boson propagator 
\begin{equation*}
d_{\alpha\beta}(k)=-g_{\alpha\beta}+\frac{\delta_{\alpha+}k_{\beta}+\delta_{\beta+}k_{\alpha}}{k^+},
\end{equation*}
 and will show that the regular, instantaneous fermion exchange as well as the instantaneous photon exchange diagrams of LFTOPT are generated by this method also. The two major points to be noted here are that (i) this method is being used to carry out the $k^{-}$-integration because of the non-vanishing arc contributions to the contour integral and (ii) it is the two-term propagator that generates the instantaneous photon diagram (alongwith the rest of the diagrams). This is due to the presence of the  interaction vertex
\begin{equation}
V_{3}=-\frac{e^{2}}{4}\int d^{2}{\bf{x}}_{\perp}dx^{-}dy^{-}(\bar{\xi}\gamma^{+}\xi)(x)|x^{-}-y^{-}|(\bar{\xi}\gamma^{+}\xi)(y)
\end{equation}
in the LF Hamiltonian \citep{mustaki}, which is otherwise  absent if we apply Lorentz condition also in the derivation of LF Hamiltonian as argued by us in Sec.  \ref{sec:propagator}. In this section, we will use the asymptotic method to prove equivalence for fermion self-energy.



The covariant expression for fermion self-energy with the two-term photon propagator is given below:
\begin{equation} \label{eq:selfen_2term}
\begin{split}
\Sigma(p)& =\frac{ie^{2}}{2m}\int\frac{d^{4}k}{(2\pi)^{4}}\frac{\gamma^{\mu}(k\llap/'+m)\gamma^{\nu}d_{\mu\nu}(k)}{[(p-k)^{2}-m^{2}+i\epsilon][k^{2}-\mu^{2}+i\epsilon]}\\
& =\Sigma_{1}^{(a)}(p)+\Sigma_{2}(p)
\end{split}
\end{equation}
where
\begin{equation}\label{eq:selfen_2term1}
\Sigma_{1}^{(a)}(p)=\frac{ie^{2}}{2m}\int\frac{d^{4}k}{(2\pi)^{4}}\frac{\gamma^{\mu}(k\llap/'_{on}+m)\gamma^{\nu}d_{\mu\nu}(k)}{[(p-k)^{2}-m^{2}+i\epsilon][k^{2}-\mu^{2}+i\epsilon]}
\end{equation}
and
\begin{equation}
\Sigma_{2}(p)=\frac{ie^{2}}{2m}\int\frac{d^{4}k}{(2\pi)^{4}}\frac{\gamma^{\mu}\gamma^{+}\gamma^{\nu}d_{\mu\nu}(k)}{2(p^{+}-k^{+})[k^{2}-\mu^{2}+i\epsilon]}
\end{equation}
as per the notations used in Sec. \ref{sec:equi_self-en}. $\bar{u}_{p,s}\Sigma_{2}(p)u_{p,s}$ has already been evaluated in that section and it was seen that it gives the expression for the instantaneous fermion exchange diagram. However, in the contour integration over $k^{-}$ in the expression for $\Sigma_{1}^{(a)}(p)$, the key observation is that, due to the presence of a factor of $k^{-}$ in $d_{\mu \nu}(k)$, there are possible arc contributions for the cases (i) $k^{-}\rightarrow{\infty}$ as $k^{+}\rightarrow{0}$ and (ii) $k^{-}\rightarrow{\infty}$ as $k^{+}\rightarrow{p^{+}}$ since in these cases, the integrand does not go to zero as $k^{-}\rightarrow\infty$. 

The Asymptotic Method consists of taking the asymptotic limits of the integrand and subtracting it from the integrand which reduces the degree of divergence. The asymptotic parts are then evaluated separately and added to the integral which can now be evaluated using the method of residues. Thus, we rewrite $\Sigma_{1}^{(a)}(p)$ as 
\[ \Sigma_{1}^{(a)}(p)= \bigg[ \Sigma_{1}^{(a)}(p) -\Sigma_{1(1)}^{(a)asy}(p) -\Sigma_{1(2)}^{(a)asy}(p) \bigg] +\Sigma_{1(1)}^{(a)asy}(p) +\Sigma_{1(2)}^{(a)asy}(p) \]
where
\[ \Sigma_{1(1)}^{(a)asy}(p) = \Lim{\substack{k^+ \rightarrow 0 \\ k^- \rightarrow \infty}} \Sigma_{1}^{(a)}(p) \]
and 
\[ \Sigma_{1(2)}^{(a)asy}(p) = \Lim{\substack{k^+ \rightarrow p^+ \\ k^- \rightarrow \infty}} \Sigma_{1}^{(a)}(p) \]
Using the identities $\gamma^{\alpha}\gamma^{\mu}\gamma^{\beta}d_{\alpha\beta}(k)=\frac{2}{k^{+}}(\gamma^{+}k^{\mu}+g^{+\mu}k\llap/)$ and $\bar{u}_{p,s}\gamma^{\mu}u_{p,s'}=2p^{\mu}\delta_{ss'}$, we see that the numerator of integrand in $\bar{u}_{p,s}\Sigma_{1}^{(a)}u_{p,s}$ is
\begin{equation}\label{eq:num_selfen}
\begin{split}
Num=&\frac{4p^{+}[{({\bf{p}}_{\perp}-{\bf{k}}_{\perp})^{2}+m^{2}}]}{(p^{+}-k^{+})}+\frac{8p^{+}(p^{+}-k^{+})k^{-}}{k^{+}}-\frac{4p^{+}({\bf{p}}_{\perp}-{\bf{k}}_{\perp}){\cdot}{\bf{k}}_{\perp}}{k^{+}}\\
&+4p^{-}(p^{+}-k^{+})-\frac{4{\bf{p}}_{\perp}{\cdot}{\bf{k}}_{\perp}(p^{+}-k^{+})}{k^{+}}-4m^{2}
\end{split}
\end{equation}
which in the asymptotic limit (i) $k^{-}\rightarrow{\infty}$ as $k^{+}\rightarrow{0}$, reduces to
\begin{equation*}
Num^{asy} _{(1)}=\frac{8p^{+}(p^{+}-k^{+})k^{-}}{k^{+}}
\end{equation*}
and the denominator of $\bar{u}_{p,s}\Sigma_{1}^{(a)}u_{p,s}$ reduces to
\begin{equation*}
Den^{asy} _{(1)}=-2k^{-}(p^{+}-k^{+})D_{1}
\end{equation*}
where $D_{1}=k^{2}-\mu^{2}+i\epsilon$.\\
Thus,
\begin{equation}
\bar{u}_{p,s'}\Sigma_{1(1)}^{(a)asy}(p)u_{p,s}=\frac{-2ie^{2}p^{+}\delta_{ss'}}{2m}\int\frac{d^{2}{\bf{k}}_{\perp}}{(2\pi)^{4}}\int\frac{dk^{+}}{(k^{+})^{2}}\int\frac{dk^{-}}{\big(k^{-}-\frac{{\bf{k}}^{2}_{\perp}+\mu^{2}-i\epsilon}{2k^{+}}\big)}
\end{equation}
The $k^{-}$-integral is evaluated in Sec. \ref{sec:equi_self-en} and is
\begin{equation*}
\int\frac{dk^{-}}{\big(k^{-}-\frac{{\bf{k}}^{2}_{\perp}+\mu^{2}-i\epsilon}{2k^{+}}\big)}=-\pi i\theta(k^{+})+\pi i\theta(-k^{+})
\end{equation*}
Hence,
\begin{equation}
\bar{u}_{p,s'}\Sigma_{1(1)}^{(a)asy}(p)u_{p,s}=\frac{e^{2}p^{+}\delta_{ss'}}{2m}\int\frac{d^{2}{\bf{k}}_{\perp}}{(2\pi)^{3}}\bigg[\int_{-\infty}^{0}\frac{dk^{+}}{(k^{+})^{2}}-\int_{0}^{\infty}\frac{dk^{+}}{(k^{+})^{2}}\bigg]
\end{equation}
Changing the variable $k^{+}$ to $-k^{+}$ in the first $k^{+}$-integral gives
\begin{equation}\label{eq:sigma2asy0}
\bar{u}_{p,s'}\Sigma_{1(1)}^{(a)asy}(p)u_{p,s}=0
\end{equation}
In the asymptotic limit (ii) $k^{-}\rightarrow{\infty}$ as $k^{+}\rightarrow{p^{+}}$, Eq.(\ref{eq:num_selfen}) reduces to
\begin{equation*}
Num^{asy} _{(2)}=\frac{8p^{+}(p^{+}-k^{+})k^{-}}{k^{+}}
\end{equation*}
and the denominator of $\bar{u}_{p,s}\Sigma_{1}^{(a)}u_{p,s}$ reduces to
\begin{equation*}
Den^{asy} _{(2)}=(2k^{+}k^{-})D_{2}
\end{equation*}
where $D_{2}=(p-k)^{2}-m^{2}+i\epsilon$.\\
So,
\begin{equation}
\bar{u}_{p,s'}\Sigma_{1(2)}^{(a)asy}(p)u_{p,s}=\frac{2ie^{2}p^{+}\delta_{ss'}}{2m}\int\frac{d^{2}{\bf{k}}_{\perp}}{(2\pi)^{4}}\int\frac{dk^{+}}{(k^{+})^{2}}\int\frac{dk^{-}}{\big(p^{-}-k^{-}-\frac{({\bf{p}}_{\perp}-{\bf{k}}_{\perp})^{2}+m^{2}-i\epsilon}{2(p^{+}-k^{+})}\big)}
\end{equation}
The above $k^{-}$-integral too is evaluated in Sec. \ref{sec:equi_self-en} and is
\begin{equation*}
\int\frac{dk^{-}}{\big(p^{-}-k^{-}-\frac{({\bf{p}}_\perp-{\bf{k}}_\perp)^{2}+m^{2}-i\epsilon}{2(p^{+}-k^{+})}\big)}=-\pi i\theta(p^{+}-k^{+})+\pi i\theta(k^{+}-p^{+})
\end{equation*}
Thus,
\begin{equation}
\bar{u}_{p,s'}\Sigma_{1(2)}^{(a)asy}(p)u_{p,s}=\frac{e^{2}p^{+}\delta_{ss'}}{2m}\int\frac{d^{2}{\bf{k}}_{\perp}}{(2\pi)^{3}}\bigg[\int_{-\infty}^{p^{+}}\frac{dk^{+}}{(k^{+})^{2}}-\int_{p^{+}}^{\infty}\frac{dk^{+}}{(k^{+})^{2}}\bigg]
\end{equation}
Changing the variable $k^{+}$ to $(p^{+}-k^{+})$ in the first $k^{+}$-integral and to $(p^{+}+k^{+})$ in the second gives
\begin{equation}\label{eq:asy_sigma}
\bar{u}_{p,s'}\Sigma_{1(2)}^{(a)asy}(p)u_{p,s} = \frac{e^{2}p^{+}\delta_{ss'}}{2m}\int\frac{d^{2}{\bf{k}}_{\perp}}{(2\pi)^{3}}\bigg[\int_{0}^{\infty}\frac{dk^{+}}{(p^{+}-k^{+})^{2}}-\int_{0}^{\infty}\frac{dk^{+}}{(p^{+}+k^{+})^{2}}\bigg]
\end{equation}
which  is  the same as the expression for  instantaneous photon exchange diagrams of Fig.(\ref{fig:selfen_inst_photon}).


Thus, we see that the asymptotic method correctly generates the instantaneous photon exchange diagrams too when the two-term gauge boson propagator is used. Now we go on to show how the regular diagram is generated. Separating the asymptotic part, we get 

\begin{equation*}
\begin{split}
\bar{u}_{p,s'}\Sigma_{1}^{(a)}(p)u_{p,s} = &\big[\bar{u}_{p,s'}\Sigma_{1}^{(a)}(p)u_{p,s} -\bar{u}_{p,s'}\Sigma_{1(1)}^{(a)asy}(p)u_{p,s}-\bar{u}_{p,s'}\Sigma_{1(2)}^{(a)asy}(p)u_{p,s}\big]\\
& +\bar{u}_{p,s'}\Sigma_{1(1)}^{(a)asy}(p)u_{p,s}+\bar{u}_{p,s'}\Sigma_{1(2)}^{(a)asy}(p)u_{p,s}\\
= &\big[\bar{u}_{p,s'}\Sigma_{1}^{(a)}(p)u_{p,s}-\bar{u}_{p,s'}\Sigma_{1(2)}^{(a)asy}(p)u_{p,s}\big]+\bar{u}_{p,s'}\Sigma_{1(2)}^{(a)asy}(p)u_{p,s}
\end{split}
\end{equation*}
since $\bar{u}_{p,s'}\Sigma_{1(1)}^{(a)asy}(p)u_{p,s}=0$ (see Eq.({\ref{eq:sigma2asy0}}))\\

It can be easily shown that  
\begin{equation}\label{eq:sigma-asy_sigma}
\bar{u}_{p,s'}\Sigma_{1}^{(a)}(p)u_{p,s}-\bar{u}_{p,s'}\Sigma_{1(2)}^{(a)asy}(p)u_{p,s}=\frac{ie^{2}}{2m}\int\frac{d^{2}k_{\perp}dk^{+}N}{(2\pi)^{4}}\int\frac{dk^{-}}{D_{1}D_{2}}
\end{equation}
where $D_{1}, D_{2}$ are defined previously and
\begin{equation*}
\begin{split}
 N= & 4p^{+}k'^{-}_{on}-\frac{4p^{+}({\bf{p}}_{\perp}-{\bf{k}}_{\perp}){\cdot}{\bf{k}}_{\perp}}{k^{+}}+4p^{-}(p^{+}-k^{+})-\frac{4{\bf{p}}_{\perp}{\cdot}{\bf{k}}_{\perp}(p^{+}-k^{+})}{k^{+}}\\
&-4m^{2}+\frac{4p^{+}(p^{+}-k^{+})({\bf{k}}^{\perp})^{2}}{(k^{+})^{2}}
\end{split}
\end{equation*}
The $k^{-}$-integral in Eq.(\ref{eq:sigma-asy_sigma}) has poles at $k^{-}_{1}=\frac{{\bf{k}}_{\perp}^{2}+\mu^{2}-i\epsilon}{2k^{+}}$ and $k^{-}_{2}=p^{-}-\frac{({\bf{p}}_{\perp}-{\bf{k}}_{\perp})^{2}+m^{2}-i\epsilon}{2(p^{+}-k^{+})}$. The integral goes to zero for $k^{+}<0$ and $k^{+}>p^{+}$ since in each of these cases, both poles lie on one side of the real axis. For $0<k^{+}<p^{+}$, $k_{1}^{-}$ lies below whereas $k_{2}^{-}$ lies above the real axis. Closing the contour below, we find
\begin{equation*}
\bar{u}_{p,s'}\Sigma_{1}^{(a)}(p)u_{p,s}-\bar{u}_{p,s'}\Sigma_{1(2)}^{(a)asy}(p)u_{p,s}=\frac{e^{2}}{m}\int\frac{d^{2}{\bf{k}}_{\perp}}{(4\pi)^{3}}\int_{0}^{p^{+}}\frac{dk^{+}N}{k^{+}(p^{+}-k^{+})[p^{-}-k^{-}_{on}-k'^{-}_{on}]}
\end{equation*}
Using the identities $\gamma^{\alpha}\gamma^{\mu}\gamma^{\beta}d_{\alpha\beta}(k)=\frac{2}{k^{+}}(\gamma^{+}k^{\mu}+g^{+\mu}k\llap/)$ and $\bar{u}_{p,s}\gamma^{\mu}u_{p,s'}=2p^{\mu}\delta_{ss'}$, it can be seen that
\begin{equation*}
\bar{u}_{p,s'}\gamma^{\mu}(k\llap/'_{on}+m)\gamma^{\nu}d_{\mu\nu}(k_{on})u_{p,s}=N
\end{equation*}
Thus,
\begin{equation}
\bar{u}_{p',s'}\Sigma_{1}u_{p,s}=\frac{e^{2}}{m}\int\frac{d^{2}{\bf{k}}_{\perp}}{(4\pi)^{3}}\int_{0}^{p^{+}}\frac{dk^{+}}{k^{+}(p^{+}-k^{+})}\frac{\bar{u}_{p,s'}\gamma^{\mu}({k\llap/}'+m)\gamma^{\nu}u_{p,s}d_{\mu\nu}(k_{on})}{p^{-}-k^{-}_{on}-k'^{-}_{on}}
\end{equation}
which is the expression for the regular diagram (Eq.(\ref{eq:se_reg})). Thus, we see that using the two-term propagator and employing the Asymptotic Method to consistently take into account the arc contributions, all the one loop self-energy diagrams in Ref.\citep{mustaki} viz. the regular diagram, the instantaneous fermion exchange diagram and the instantaneous photon exchange diagrams are generated. As shown in Sec. \ref{sec:equi_self-en}, if one uses the three-term gauge boson propagator, there is an extra contribution due to the third term of the propagator, which will cancel Eq.(\ref{eq:asy_sigma}) and thus in this method also, the three-term propagator generates only the regular and instantaneous fermion exchange diagrams. This is consistent with the arguments presented  in Sec. \ref{sec:propagator}.

\section{One Loop Vertex Correction in LFTOPT}\label{sec:equi_vertcorr}



\begin{figure}[h!]
\centering
\subfloat[]
{\includegraphics[scale=0.65]{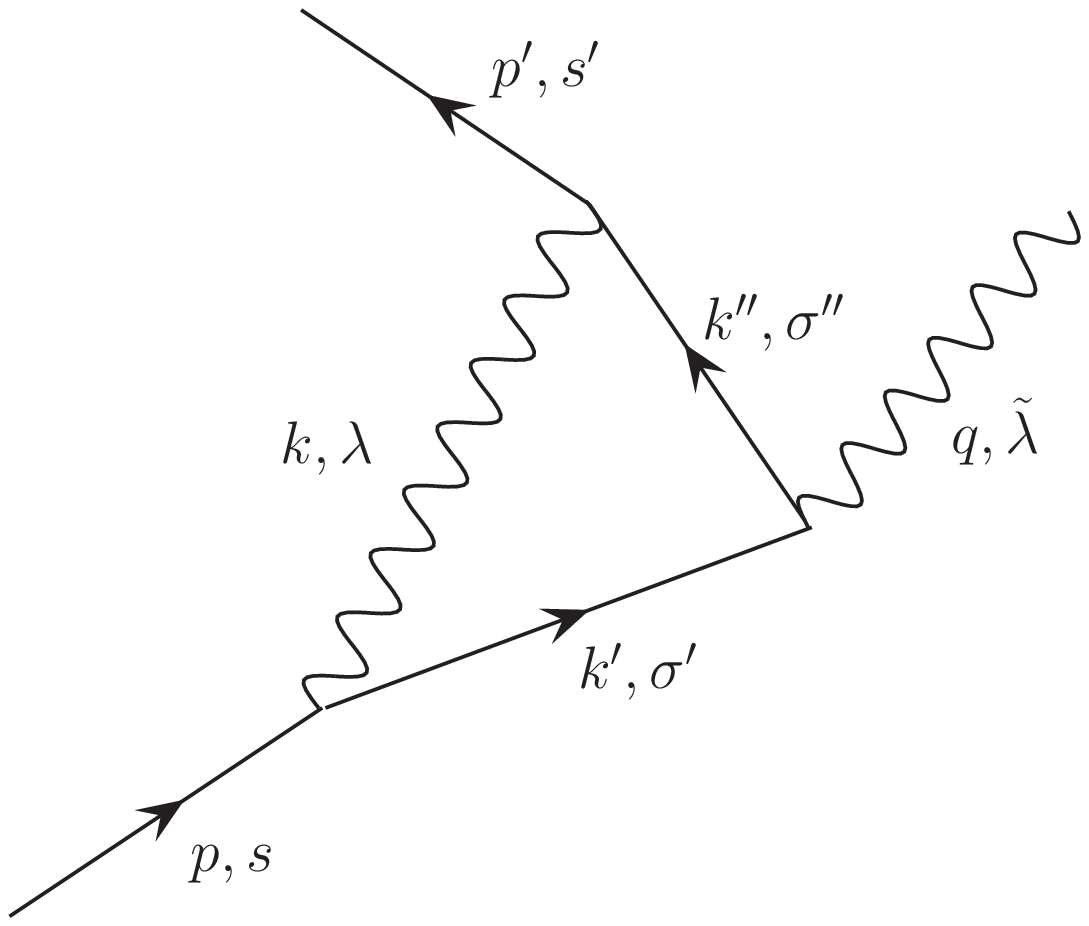}}
\hspace{1cm}
\subfloat[]
{\includegraphics[scale=0.57]{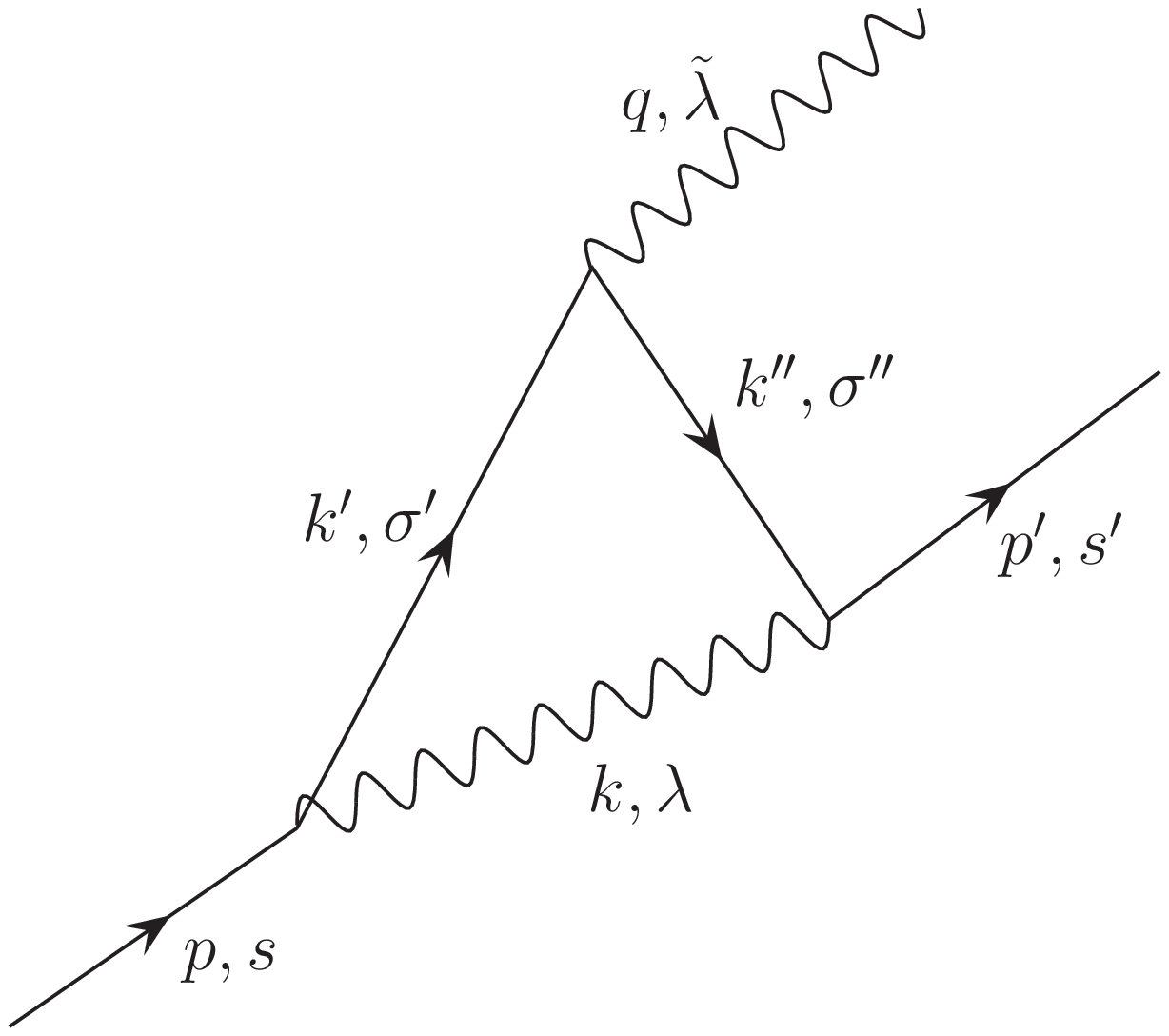}}
\qquad
\subfloat[]
{\includegraphics[scale=0.6]{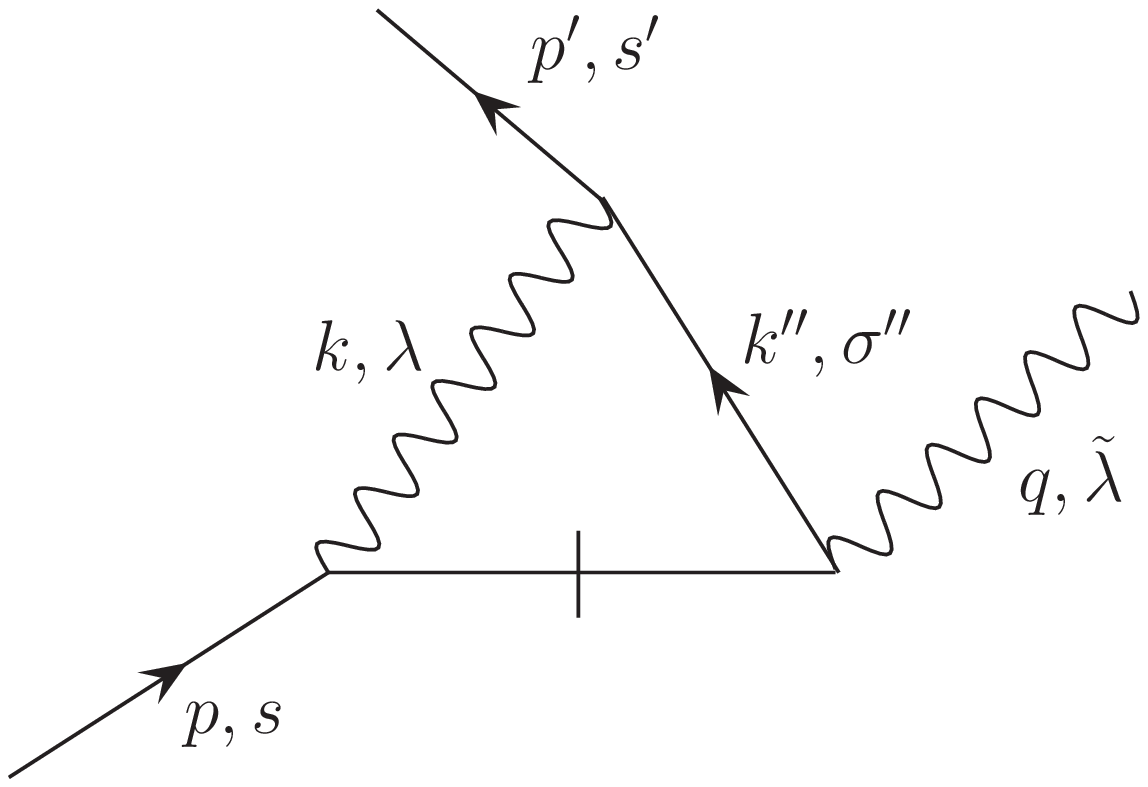}}
\hspace{1cm}
\subfloat[]
{\includegraphics[scale=0.6]{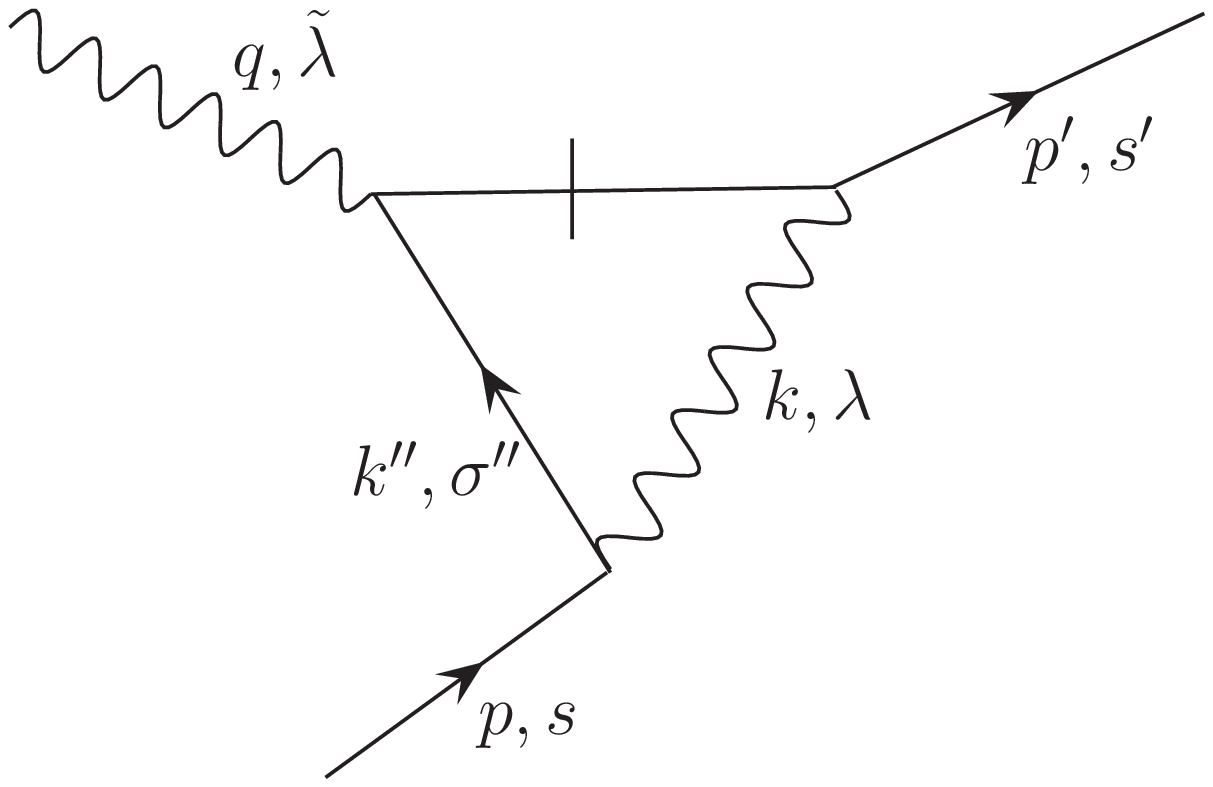}}
\caption{"Regular" and instantaneous fermion exchange vertex correction diagrams}
\label{fig:reg_inst_photon}
\end{figure}

One loop renormalization of LFQED has been discussed in detail in Ref.\cite{mustaki}, where the authors have enlisted all the one loop diagrams contributing to $\Lambda^\mu$. We have presented, in Fig.\ref{fig:reg_inst_photon}, all the connected diagrams that contribute to the process \cite{mustaki, swati}. The rest of the diagrams for vertex correction given in Ref.\citep{mustaki} are corrections to external legs and hence can be absorbed in  renormalization constants. Thus, the only diagrams relevant here are those given in Fig.\ref{fig:reg_inst_photon}. 
Figs.\ref{fig:reg_inst_photon}(a) and (b), which we call the regular diagrams, contain only the standard QED vertex. These two have been evaluated for the `$+$' component in Ref.\cite{mustaki} using LFTOPT. Diagrams in Figs.\ref{fig:reg_inst_photon}(c) and (d) contain the instantaneous fermion vertex and were not evaluated by Mustaki {\it et al.} and by us \cite{swati} as these two diagrams do not contribute to $\Lambda^+$ because of the structure of $\gamma$-matrices. Both the works discussed the evaluation and equivalence of the `$+$' component only. Here, we extend this study to a general $\Lambda^{\mu}$.

Contributions of the regular diagrams in Figs.\ref{fig:reg_inst_photon}(a) and (b) are given by 
\begin{equation} \label{eq:vc_reg_diag1}
 \begin{split}
 \Lambda^{\mu}_{(a)} = &\lambda \int_{-\infty}^{+\infty}\frac{d^{2}{\bf{k}}_{\perp}}{(4\pi)^{3}}\int_{0}^{p^{+}-q^{+}}\frac{dk^{+}}{k^{+}k'^{+}k''^{+}}\frac{\gamma^{\alpha}({k\llap/}''+m)\gamma^{\mu}({k\llap/}'+m)\gamma^{\beta}d_{\alpha\beta}(k)}{(p^{-}-k^{-}-k'^{-})(p^{-}-q^{-}-k^{-}-k''^{-})}
 \end{split}
\end{equation}
and 
\begin{equation} \label{eq:vc_reg_diag2}
 \begin{split}
 \Lambda^{\mu}_{(b)} = &-\lambda \int_{-\infty}^{+\infty}\frac{d^{2}{\bf{k}}_{\perp}}{(4\pi)^{3}}\int_{p^{+}-q^{+}}^{p^{+}}\frac{dk^{+}}{k^{+}k'^{+}k''^{+}}\frac{\gamma^{\alpha}({k\llap/}''+m)\gamma^{\mu}({k\llap/}'+m)\gamma^{\beta}d_{\alpha\beta}(k)}{(p^{-}-k^{-}-k'^{-})(p^{-}-p'^{-}-k'^{-}+k''^{-})}
 \end{split}
\end{equation}
respectively
where $\lambda^{-1}=(2\pi)^{3/2}\sqrt{2p^{+}}\sqrt{2p'^{+}}\sqrt{2q^{+}}$.
As mentioned earlier, the diagrams in Figs.\ref{fig:reg_inst_photon}(c) and (d) have not been evaluated earlier and hence we present the calculation of these in detail below.

In perturbation theory, the transition amplitude has the expansion
\begin{equation}
T=V+V\frac{1}{p^{-}-H_{0}}V+...
\end{equation}
For the diagram of Fig.\ref{fig:reg_inst_photon}(c), the transition amplitude upto order $e^{3}$ is written as\\$T^{(c)}_{p,p',q}=e^{3}\bar{u}_{p',s'}\Lambda^{\mu}_{(c)}u_{p,s}\epsilon^{\tilde\lambda}_{k}(q)\delta^{3}[{\bf{p}}-({\bf{p}}'+{\bf{q}})]\theta(p^{+})\theta(p^{+}-q^{+})$\\
whereas for the diagram of Fig.\ref{fig:reg_inst_photon}(d), we write\\
$ T^{(d)}_{p,p',q} = e^{3}\bar{u}_{p',s'}\Lambda^{\mu}_{(d)}u_{p,s}\epsilon^{\tilde\lambda}_{j}(q)\delta^{3}[{\bf{p}}-({\bf{p}}'+{\bf{q}})]\theta(p^{+})\theta(p^{+}-q^{+})$.\\
\noindent The transition amplitudes due to the instantaneous fermion exchange diagrams are  obtained, following the standard procedure,  by inserting complete sets of states which leads to the following expressions (Details presented in Appendix \ref{app:details_sec_III}):
\begin{equation*}
 \begin{split}
  T^{(c)}_{p,p',q} = &\mel{p',s';q,\tilde{\lambda}}{V_1\frac{1}{p^{-}-H_0}V_2}{p,s}\\
  =&\int d^{3}{\bf{k}}''d^{3}{\bf{k}}d^{3}{\bf{k}}''_1d^{3}{\bf{k}}_1\theta(k''^+)\theta(k^+)\theta(k''^{+}_1)\theta(k^{+}_1)\sum_{\sigma'',\lambda,\sigma''_1,\lambda_1}\mel{p',s';q,\tilde{\lambda}}{V_1}{k'',\sigma'';k,\lambda;q,\tilde{\lambda}}\\
  &\mel{k'',\sigma'';k,\lambda;q,\tilde{\lambda}}{\frac{1}{p^{-}-H_0}}{k''_1,\sigma''_1;k_1,\lambda_1;q,\tilde{\lambda}}\mel{k''_1,\sigma''_1;k_1,\lambda_1;q,\tilde{\lambda}}{V_2}{p,s}\\
  =&\int\frac{d^{3}{\bf{k}}''d^{3}{\bf{k}}\theta(k''^+)\theta(k^+)}{p^{-}-k''^{-}-k^{-}-q^{-}}\sum_{\sigma'',\lambda}\mel{p',s';q,\tilde{\lambda}}{V_1}{k'',\sigma'';k,\lambda;q,\tilde{\lambda}}\mel{k'',\sigma'';k,\lambda;q,\tilde{\lambda}}{V_2}{p,s}
 \end{split}
\end{equation*}
for the diagram in  Fig.\ref{fig:reg_inst_photon}(c)  and
\begin{equation*}
 \begin{split}
  T^{(d)} _{p,p',q} = &\mel{p',s';q,\tilde{\lambda}}{V_2\frac{1}{p^{-}-H_0}V_1}{p,s}\\
  =&\int d^{3}{\bf{k}}''d^{3}{\bf{k}}d^{3}{\bf{k}}''_1d^{3}{\bf{k}}_1\theta(k''^+)\theta(k^+)\theta(k''^{+}_1)\theta(k^{+}_1)\sum_{\sigma'',\lambda,\sigma''_1,\lambda_1}\mel{p',s';q,\tilde{\lambda}}{V_2}{k'',\sigma'';k,\lambda}\\
  &\mel{k'',\sigma'';k,\lambda}{\frac{1}{p^{-}-H_0}}{k''_1,\sigma''_1;k_1,\lambda_1}\mel{k''_1,\sigma''_1;k_1,\lambda_1}{V_1}{p,s}\\
  =&\int\frac{d^{3}{\bf{k}}''d^{3}{\bf{k}}\theta(k''^+)\theta(k^+)}{p^{-}-k''^{-}-k^{-}}\sum_{\sigma'',\lambda}\mel{p',s';q,\tilde{\lambda}}{V_2}{k'',\sigma'';k,\lambda}\mel{k'',\sigma'';k,\lambda}{V_1}{p,s}
 \end{split}
\end{equation*}
for the diagram in Fig.\ref{fig:reg_inst_photon}(d), where $d^{3}{\bf{k}}=dk^{+}d^{2}{\bf{k}}_\perp$.
Calculating each of these matrix elements by substituting the appropriate term  of the interaction Hamiltonian and expanding the  fields in their Fourier components, one obtains the following expressions for the transition amplitudes

\begin{equation} \label{eq:amp1}
 \begin{split}
 T^{(c)}_{p,p',q} = &e^{3}\bar{u}_{p',s'}\bigg[\lambda\int\frac{d^{3}{\bf{k}}\theta(k^{+})\theta(p'^{+}-k^{+})}{(4\pi)^{3}k^{+}k''^{+}(p^{+}-k^{+})}\frac{\gamma^{\alpha}({k\llap/}''+m)\gamma^{k}\gamma^{+}\gamma^{j}d_{\alpha j}(k)}{(p^{-}-k''^{-}-k^{-}-q^{-})}\bigg]\\
 &u_{p,s}\epsilon^{\tilde\lambda}_{k}(q)\delta^{3}[{\bf{p}}-({\bf{p}}'+{\bf{q}})]\theta(p^{+})\theta(p^{+}-q^{+})
 \end{split}
\end{equation}
and
\begin{equation} \label{eq:amp2}
 \begin{split}
 T^{(d)}_{p,p',q} = &e^{3}\bar{u}_{p',s'}\bigg[\lambda\int\frac{d^{3}{\bf{k}}\theta(k^{+})\theta(p^{+}-k^{+})}{(4\pi)^{3}k^{+}k'^{+}(p^{+}-k^{+}-q^{+})}\frac{\gamma^{k}\gamma^{+}\gamma^{j}({k\llap/}'+m)\gamma^{\beta}d_{k\beta}(k)}{(p^{-}-k'^{-}-k^{-})}\bigg]\\
 &u_{p,s}\epsilon^{\tilde\lambda}_{j}(q)\delta^{3}[{\bf{p}}-({\bf{p}}'+{\bf{q}})]\theta(p^{+})\theta(p^{+}-q^{+})
 \end{split}
\end{equation}
respectively for the two diagrams. Using Eq.(\ref{eq:appendix_b4}), we obtain 
\begin{equation}\label{eq:vc_amp_2a}
 \begin{split}
 T^{(c)}_{p,p',q} = &e^{3}\bar{u}_{p',s'}\bigg[\lambda \int_{-\infty}^{+\infty}\frac{d^{2}{\bf{k}}_{\perp}}{(4\pi)^{3}}\int_{0}^{p'{+}}\frac{dk^{+}}{k^{+}k'^{+}k''^{+}}\frac{\gamma^{\alpha}({k\llap/}''+m)\gamma^{\mu}\gamma^{+}\gamma^{\beta}d_{\alpha\beta}(k)}{(p'^{-}-k^{-}-k''^{-})}\bigg]\\
 &u_{p,s}\epsilon^{\tilde\lambda}_{\mu}(q)\delta^{3}[{\bf{p}}-({\bf{p}}'+{\bf{q}})]\theta(p^{+})\theta(p^{+}-q^{+})
 \end{split}
\end{equation}
and hence,
\begin{equation} \label{eq:vc_inst_ferm1}
 \begin{split}
 \Lambda^{\mu}_{(c)} = &\lambda \int_{-\infty}^{+\infty}\frac{d^{2}{\bf{k}}_{\perp}}{(4\pi)^{3}}\int_{0}^{p'{+}}\frac{dk^{+}}{k^{+}k'^{+}k''^{+}}\frac{\gamma^{\alpha}({k\llap/}''+m)\gamma^{\mu}\gamma^{+}\gamma^{\beta}d_{\alpha\beta}(k)}{(p'^{-}-k^{-}-k''^{-})}
 \end{split}
\end{equation}
In a similar fashion, using Eq.(\ref{eq:appendix_b5}), we obtain
\begin{equation}\label{eq:vc_amp_2b}
 \begin{split}
 T^{(d)}_{p,p',q} = &e^{3}\bar{u}_{p',s'}\bigg[\lambda \int_{-\infty}^{+\infty}\frac{d^{2}{\bf{k}}_{\perp}}{(4\pi)^{3}}\int_{0}^{p{+}}\frac{dk^{+}}{k^{+}k'^{+}k''^{+}}\frac{\gamma^{\alpha}\gamma^{+}\gamma^{\mu}({k\llap/}'+m)\gamma^{\beta}d_{\alpha\beta}(k)}{(p^{-}-k^{-}-k'^{-})}\bigg]\\
 &u_{p,s}\epsilon^{\tilde\lambda}_{\mu}(q)\delta^{3}[{\bf{p}}-({\bf{p}}'+{\bf{q}})]\theta(p^{+})\theta(p^{+}-q^{+})
 \end{split}
\end{equation}
and hence,
\begin{equation} \label{eq:vc_inst_ferm2}
 \begin{split}
 \Lambda^{\mu}_{(d)} = &\lambda \int_{-\infty}^{+\infty}\frac{d^{2}{\bf{k}}_{\perp}}{(4\pi)^{3}}\int_{0}^{p{+}}\frac{dk^{+}}{k^{+}k'^{+}k''^{+}}\frac{\gamma^{\alpha}\gamma^{+}\gamma^{\mu}({k\llap/}'+m)\gamma^{\beta}d_{\alpha\beta}(k)}{(p^{-}-k^{-}-k'^{-})}
 \end{split}
\end{equation}

\section{Equivalence of Covariant and Light-Front Expressions of one loop Vertex Correction}\label{sec:equi_vertcorrCov}

In this section, we present the proof of equivalence for the one loop vertex correction. In Ref.\cite{swati}, we established the equivalence of covariant and LF expressions for $\Lambda^+$ i.e. the `$+$' component of the one loop vertex correction $\Lambda^\mu$. Here, we present a more general proof valid for all components of $\Lambda^\mu$.

\par
The standard covariant expression for vertex correction in the light-front gauge comes from Fig.\ref{fig:reg_inst_photon}(a) which is the only diagram that contributes to $\Lambda^{\mu}$ in covariant theory. It is given by
\begin{equation} \label{eq:vc_main_3term}
\Lambda^{\mu}(p,p^{\prime},q) = ie^3\int \frac {d^{4}k}{(2\pi)^4}\frac{\gamma^{\alpha}({p\llap/^{\prime}}-{k\llap/}+m)\gamma^{\mu}({p\llap/}-{k\llap/}+m)\gamma^{\beta}d^{\prime}_{\alpha\beta}(k)}{[(p-k)^2-m^2+i\epsilon][(p^{\prime}-k)^2-m^2+i\epsilon][k^2-{\mu}^2+i\epsilon]}
\end{equation}
where we have used the three-term photon propagator 
\begin{equation*}
 \begin{split}
 d^{\prime}_{\alpha\beta}(k)=&d_{\alpha\beta}(k)-\frac{\delta_{\alpha+}\delta_{\beta+}k^2}{(k^{+})^2}\\
 =&-g_{\alpha\beta}+\frac{\delta_{\alpha+}k_{\beta}+\delta_{\beta+}k_{\alpha}}{k^+}-\frac{\delta_{\alpha+}\delta_{\beta+}k^2}{(k^{+})^2}
 \end{split}
\end{equation*}


In order to show that this standard covariant expression for vertex correction is equivalent to the expressions calculated in the light-front time-ordered perturbation theory diagrams given in  Fig.\ref{fig:reg_inst_photon}, we split the fermion momenta into on-shell and off-shell parts as was done in the case of fermion self-energy and vacuum polarization. 
Similar to Eq.(\ref{eq:splitting_1}), $p\llap/'-k\llap/$ can be written as
\begin{equation}\label{eq:splitting_2}
p\llap/'-k\llap/=k\llap/''_{on}+\frac{\gamma^{+}[(p'-k)^{2}-m^{2}]}{2(p'^{+}-k^{+})}
\end{equation}
Using Eqs.(\ref{eq:splitting_1}) and (\ref{eq:splitting_2}), Eq.(\ref{eq:vc_main_3term}) becomes
\begin{equation} \label{eq:vc_splitup}
\begin{split}
\Lambda^{\mu}(p,p^{\prime},q) =& ie^3\int \frac {d^{4}k}{(2\pi)^4}\frac{\gamma^{\alpha}({k\llap/''_{on}}+m)\gamma^{\mu}({k\llap/'_{on}}+m)\gamma^{\beta}d^{\prime}_{\alpha\beta}(k)}{[(p-k)^2-m^2+i\epsilon][(p^{\prime}-k)^2-m^2+i\epsilon][k^2-{\mu}^2+i\epsilon]}\\
+& ie^3\int \frac {d^{4}k}{(2\pi)^4}\frac{\gamma^{\alpha}({k\llap/''_{on}}+m)\gamma^{\mu}\gamma^{+}\gamma^{\beta}d^{\prime}_{\alpha\beta}(k)}{2(p^{+}-k^{+})[(p^{\prime}-k)^2-m^2+i\epsilon][k^2-{\mu}^2+i\epsilon]}\\
+& ie^3\int \frac {d^{4}k}{(2\pi)^4}\frac{\gamma^{\alpha}\gamma^{+}\gamma^{\mu}({k\llap/'_{on}}+m)\gamma^{\beta}d^{\prime}_{\alpha\beta}(k)}{2(p'^{+}-k^{+})[(p-k)^2-m^2+i\epsilon][k^2-{\mu}^2+i\epsilon]}\\
+& ie^3\int \frac {d^{4}k}{(2\pi)^4}\frac{\gamma^{\alpha}\gamma^{+}\gamma^{\mu}\gamma^{+}\gamma^{\beta}d^{\prime}_{\alpha\beta}(k)}{2(p^{+}-k^{+})2(p'^{+}-k^{+})[k^2-{\mu}^2+i\epsilon]}
\end{split}
\end{equation}
The last integral in the above equation does not contribute to the transition amplitude $T_{p,p',q}=\bar{u}_{p,s'}\Lambda^{\mu}_{p,p',q}u_{p,s}\epsilon^{\tilde{\lambda}}_{\mu}(q)$ as can be seen using $\epsilon^{\tilde{\lambda}}_{-}=0$, $(\gamma^{+})^2=0$ and the anticommutation relations of $\gamma$-matrices. The  identity $(\gamma^{+})^2=0$ also leads to the fact that the third term of the photon propagator viz. $-\frac{\delta_{\alpha+}\delta_{\beta+}k^2}{(k^{+})^2}$ provides null contributions to the second and third integrals of the above equation. Hence, Eq.(\ref{eq:vc_splitup}) reduces to
\begin{equation}\label{eq:vc_split}
\Lambda^{\mu}_{p,p',q}=\Lambda^{\mu}_{1_{p,p',q}}+\Lambda^{\mu}_{2_{p,p',q}}+\Lambda^{\mu}_{3_{p,p',q}}+\Lambda^{\mu}_{4_{p,p',q}}
\end{equation}
where
\begin{equation}\label{eq:LambdaMu1}
\Lambda^{\mu}_{1_{p,p',q}}=ie^3\int \frac {d^{4}k}{(2\pi)^4}\frac{\gamma^{\alpha}({k\llap/''_{on}}+m)\gamma^{\mu}({k\llap/'_{on}}+m)\gamma^{\beta}d_{\alpha\beta}(k)}{[(p-k)^2-m^2+i\epsilon][(p^{\prime}-k)^2-m^2+i\epsilon][k^2-{\mu}^2+i\epsilon]},
\end{equation}
\begin{equation}\label{eq:LambdaMu2}
\Lambda^{\mu}_{2_{p,p',q}}=-ie^3\int \frac {d^{4}k}{(2\pi)^4}\frac{\gamma^{+}({k\llap/''_{on}}+m)\gamma^{\mu}({k\llap/'_{on}}+m)\gamma^{+}}{(k^{+})^2[(p-k)^2-m^2+i\epsilon][(p^{\prime}-k)^2-m^2+i\epsilon]},
\end{equation}
\begin{equation}\label{eq:LambdaMu3}
\Lambda^{\mu}_{3_{p,p',q}}=ie^3\int \frac {d^{4}k}{(2\pi)^4}\frac{\gamma^{\alpha}({k\llap/''_{on}}+m)\gamma^{\mu}\gamma^{+}\gamma^{\beta}d_{\alpha\beta}(k)}{2(p^{+}-k^{+})[(p^{\prime}-k)^2-m^2+i\epsilon][k^2-{\mu}^2+i\epsilon]},
\end{equation}
\begin{equation}\label{eq:LambdaMu4}
\Lambda^{\mu}_{4_{p,p',q}}=ie^3\int \frac {d^{4}k}{(2\pi)^4}\frac{\gamma^{\alpha}\gamma^{+}\gamma^{\mu}({k\llap/'_{on}}+m)\gamma^{\beta}d_{\alpha\beta}(k)}{2(p'^{+}-k^{+})[(p-k)^2-m^2+i\epsilon][k^2-{\mu}^2+i\epsilon]}
\end{equation}
There are sufficient powers of $k^{-}$ in the denominators of $\Lambda^{\mu}_{1_{p,p',q}}$ and $\Lambda^{\mu}_{2_{p,p',q}}$ to make the $k^{-}$-integral vanish  on the arc at infinity and hence, there are no arc contributions in the case of $\Lambda^{\mu}_{1_{p,p',q}}$ and $\Lambda^{\mu}_{2_{p,p',q}}$. Thus, a naive contour integration using the method of residues gives the required result. The integrals are explicitly evaluated below.

$\Lambda^{\mu}_{1_{p,p',q}}$ can be written as
\begin{equation*}
\Lambda^{\mu}_{1_{p,p',q}}=ie^3\int \frac {d^{2}{\bf{k}}_{\perp}dk^{+}}{(2\pi)^4}\frac{\gamma^{\alpha}({k\llap/''_{on}}+m)\gamma^{\mu}({k\llap/'_{on}}+m)\gamma^{\beta}}{2k^{+}2(p^{+}-k^{+})2(p'^{+}-k^{+})}\ I_{1}
\end{equation*}
where
\begin{equation*}
I_{1}=\int\frac{dk^{-}d_{\alpha\beta}(k)}{\bigg[k^{-}-\bigg[\frac{{\bf{k}}_{\perp}^{2}+\mu^{2}-i\epsilon}{2k^{+}}\bigg]\bigg]\bigg[p^{-}-k^{-}-\bigg[\frac{({\bf{p}}_{\perp}-{\bf{k}}_{\perp})^{2}+m^{2}-i\epsilon}{2(p^{+}-k^{+})}\bigg]\bigg]\bigg[p'^{-}-k^{-}-\bigg[\frac{({\bf{p}}'_{\perp}-{\bf{k}}_{\perp})^{2}+m^{2}-i\epsilon}{2(p'^{+}-k^{+})}\bigg]\bigg]}
\end{equation*}
which has poles at $k^{-}_{1}=\frac{{\bf{k}}_{\perp}^{2}+\mu^{2}-i\epsilon}{2k^{+}}$, $k^{-}_{2}=p^{-}-\frac{({\bf{p}}_{\perp}-{\bf{k}}_{\perp})^{2}+m^{2}-i\epsilon}{2(p^{+}-k^{+})}$ and $k^{-}_{3}=p'^{-}-\frac{({\bf{p}}'_{\perp}-{\bf{k}}_{\perp})^{2}+m^{2}-i\epsilon}{2(p'^{+}-k^{+})}$.
For $k^{+}<0$, all three poles lie above the real axis. Thus, by closing the contour in the lower half-plane, the integral vanishes. Similarly, for $k^{+}>p^{+}$, since all three poles lie below the real axis, the integral vanishes on closing the contour in the upper half-plane. For $0<k^{+}<p'^{+}$, we close the contour below the real axis. $k^{-}_{2}$ and $k^{-}_{3}$ do not contribute as they fall outside the contour. The only contribution to $I_{1}$ for $0<k^{+}<p'^{+}$ comes from pole at $k^{-}_{1}$ and using the residue theorem one obtains
\begin{equation*}
I_1 = \frac{-2\pi i d_{\alpha\beta}(k_{on})\theta(k^{+})\theta(p'^{+}-k^{+})}{\bigg[p^{-}-\bigg[\frac{{\bf{k}}_{\perp}^{2}+\mu^{2}-i\epsilon}{2k^{+}}\bigg]-\bigg[\frac{({\bf{p}}_{\perp}-{\bf{k}}_{\perp})^{2}+m^{2}-i\epsilon}{2(p^{+}-k^{+})}\bigg]\bigg]\bigg[p'^{-}-\bigg[\frac{{\bf{k}}_{\perp}^{2}+\mu^{2}-i\epsilon}{2k^{+}}\bigg]-\bigg[\frac{({\bf{p}}'_{\perp}-{\bf{k}}_{\perp})^{2}+m^{2}-i\epsilon}{2(p'^{+}-k^{+})}\bigg]\bigg]}
\end{equation*}
In the case where $p'^{+}<k^{+}<p^{+}$, only $k^{-}_{2}$ contributes to $I_{1}$ on closing the contour above the real axis since $k^{-}_{1}$ and $k^{-}_{3}$ lie below the real axis. This contribution is equal to 
 \begin{equation*}
 I_1= \frac{-2\pi i d_{\alpha\beta}(k^{+}, k_{2}^{-}, k^{\perp})\theta(k^{+}-p'^{+})\theta(p^{+}-k^{+})}{\bigg[p^{-}-\bigg[\frac{({\bf{p}}_{\perp}-{\bf{k}}_{\perp})^{2}+m^{2}-i\epsilon}{2(p^{+}-k^{+})}\bigg]-\bigg[\frac{{\bf{k}}_{\perp}^{2}+\mu^{2}-i\epsilon}{2k^{+}}\bigg]\bigg]\bigg[p'^{-}-p^{-}+\bigg[\frac{({\bf{p}}_{\perp}-{\bf{k}}_{\perp})^{2}+m^{2}-i\epsilon}{2(p^{+}-k^{+})}\bigg]-\bigg[\frac{({\bf{p}}'_{\perp}-{\bf{k}}_{\perp})^{2}+m^{2}-i\epsilon}{2(p'^{+}-k^{+})}\bigg]\bigg]}
 \end{equation*}
Thus,
\begin{equation} \label{eq:vc_lambda1}
\begin{split}
\Lambda^{\mu}_{1_{p,p',q}}=&e^3\int \frac {d^{2}{\bf{k}}_{\perp}}{(4\pi)^3}\int_{0}^{p'^{+}}\frac{dk^{+}}{k^{+}k'^{+}k''^{+}}\frac{\gamma^{\alpha}({k\llap/''_{on}}+m)\gamma^{\mu}({k\llap/'_{on}}+m)\gamma^{\beta}d_{\alpha\beta}(k_{on})}{(p^{-}-k^{-}_{on}-k'^{-}_{on})(p^{-}-q^{-}-k^{-}_{on}-k''^{-}_{on})}\\
-&e^3\int \frac {d^{2}{\bf{k}}_{\perp}}{(4\pi)^3}\int_{p'^{+}}^{p^{+}}\frac{dk^{+}}{k^{+}k'^{+}k''^{+}}\frac{\gamma^{\alpha}({k\llap/''_{on}}+m)\gamma^{\mu}({k\llap/'_{on}}+m)\gamma^{\beta}d_{\alpha\beta}(k^{+}, k_{2}^{-}, k^{\perp})}{(p^{-}-k^{-}_{on}-k'^{-}_{on})(p^{-}-p'^{-}-k'^{-}_{on}+k''^{-}_{on})}
\end{split}
\end{equation}
However,
\begin{equation}
d_{\alpha\beta}(k_{2}^{-})=d_{\alpha\beta}(k_{on}^{-})+\frac{2(p^{-}-{k'}_{on}^{-}-k_{on}^{-})\delta_{\alpha+}\delta_{\beta+}}{k^{+}}
\end{equation}
which leads to 
\begin{equation} \label{eq:vc_lambda1}
\begin{split}
\Lambda^{\mu}_{1_{p,p',q}}=&e^3\int \frac {d^{2}{\bf{k}}_{\perp}}{(4\pi)^3}\int_{0}^{p'^{+}}\frac{dk^{+}}{k^{+}k'^{+}k''^{+}}\frac{\gamma^{\alpha}({k\llap/''_{on}}+m)\gamma^{\mu}({k\llap/'_{on}}+m)\gamma^{\beta}d_{\alpha\beta}(k_{on})}{(p^{-}-k^{-}_{on}-k'^{-}_{on})(p^{-}-q^{-}-k^{-}_{on}-k''^{-}_{on})}\\
-&e^3\int \frac {d^{2}{\bf{k}}_{\perp}}{(4\pi)^3}\int_{p'^{+}}^{p^{+}}\frac{dk^{+}}{k^{+}k'^{+}k''^{+}}\frac{\gamma^{\alpha}({k\llap/''_{on}}+m)\gamma^{\mu}({k\llap/'_{on}}+m)\gamma^{\beta}d_{\alpha\beta}(k_{on})}{(p^{-}-k^{-}_{on}-k'^{-}_{on})(p^{-}-p'^{-}-k'^{-}_{on}+k''^{-}_{on})}\\
-&2e^3\int \frac {d^{2}{\bf{k}}_{\perp}}{(4\pi)^3}\int_{p'^{+}}^{p^{+}}\frac{dk^{+}}{(k^{+})^{2}k'^{+}k''^{+}}\frac{\gamma^{+}({k\llap/''_{on}}+m)\gamma^{\mu}({k\llap/'_{on}}+m)\gamma^{+}}{(p^{-}-p'^{-}-k'^{-}_{on}+k''^{-}_{on})}
\end{split}
\end{equation}

Next, we consider Eq.(\ref{eq:LambdaMu2}), which can be written as 
\begin{equation*}
\Lambda^{\mu}_{2_{p,p',q}}=-ie^3\int \frac {d^{2}{\bf{k}}_{\perp}dk^{+}}{(2\pi)^4}\frac{\gamma^{+}({k\llap/''_{on}}+m)\gamma^{\mu}({k\llap/'_{on}}+m)\gamma^{+}}{{(k^{+})^2}2(p^{+}-k^{+})2(p'^{+}-k^{+})}\ I_{2}
\end{equation*}
where
\begin{equation*}
I_{2}=\int\frac{dk^{-}}{\bigg[p^{-}-k^{-}-\bigg[\frac{({\bf{p}}_{\perp}-{\bf{k}}_{\perp})^{2}+m^{2}-i\epsilon}{2(p^{+}-k^{+})}\bigg]\bigg]\bigg[p'^{-}-k^{-}-\bigg[\frac{({\bf{p}}'_{\perp}-{\bf{k}}_{\perp})^{2}+m^{2}-i\epsilon}{2(p'^{+}-k^{+})}\bigg]\bigg]}.
\end{equation*}
$I_{2}$ has a pole at $k^{-}_{1}=p^{-}-\frac{({\bf{p}}_{\perp}-{\bf{k}}_{\perp})^{2}+m^{2}-i\epsilon}{2(p^{+}-k^{+})}$ and at $k^{-}_{2}=p'^{-}-\frac{({\bf{p}}'_{\perp}-{\bf{k}}_{\perp})^{2}+m^{2}-i\epsilon}{2(p'^{+}-k^{+})}$. For $k^{+}<p'^{+}$, both the poles lie above the real axis and the integral vanishes on closing the contour below it whereas for $k^{+}>p^{+}$, they lie below the real axis and hence the integral goes to zero when the contour is closed above. For $p^{+}>k^{+}>p'^{+}$, we close the contour below the real axis. Thus, the only contribution to $I_{2}$ comes from the residue at $k^{-}_{2}$ and is equal to 
\begin{equation*}
I_{2}=\frac{2\pi i \theta(k^{+}-{p'}^{+})\theta({p}^{+}-k^{+})}{\bigg[p^{-}-p'^{-}-\bigg[\frac{({\bf{p}}_{\perp}-{\bf{k}}_{\perp})^{2}+m^{2}-i\epsilon}{2(p^{+}-k^{+})}\bigg]+\bigg[\frac{({\bf{p}}'_{\perp}-{\bf{k}}_{\perp})^{2}+m^{2}-i\epsilon}{2(p'^{+}-k^{+})}\bigg]\bigg]}
\end{equation*} 
Thus,
\begin{equation} \label{eq:vc_lambda2}
\Lambda^{\mu}_{2_{p,p',q}}=2e^3\int \frac {d^{2}{\bf{k}}_{\perp}}{(4\pi)^3}\int_{p'^{+}}^{p^{+}}\frac{dk^{+}}{(k^{+})^{2}k'^{+}k''^{+}}\frac{\gamma^{+}({k\llap/''_{on}}+m)\gamma^{\mu}({k\llap/'_{on}}+m)\gamma^{+}}{(p^{-}-p'^{-}-k'^{-}_{on}+k''^{-}_{on})}
\end{equation}
It is to be noted that the last term of Eq.(\ref{eq:vc_lambda1}) is cancelled by $\Lambda^{\mu}_{2_{p,p',q}}$, which actually  has arisen from the third term of the photon propagator.

The numerator of the integrand in $\Lambda^{\mu}_{3_{p,p',q}}$ of Eq.(\ref{eq:LambdaMu3}) can be written as
\begin{equation}
\begin{split}
\gamma^{\alpha}({k\llap/''_{on}}+m)\gamma^{\mu}\gamma^{+}\gamma^{\beta}d_{\alpha\beta}(k)=&2\gamma^{+}\gamma^{\mu}k\llap/''_{on}+k''^{+}\big(2\gamma^{\mu}\gamma^{+}\gamma^{-}-\frac{2\gamma^{\mu}\gamma^{+}\gamma^{\perp}{\cdot}{\bf{k}}^{\perp}}{k^{+}}\big)-4mg^{\mu+}\\
&-2g^{+\mu}(k\llap/''_{on}-m)\big(\gamma^{+}\gamma^{-}-\frac{\gamma^{+}\gamma^{\perp}{\cdot}{\bf{k}}^{\perp}}{k^{+}}\big)
\end{split}
\end{equation}
using the identities
\begin{equation}
\gamma^{\alpha}\gamma^{\mu}\gamma^{\nu}\gamma^{\beta}d_{\alpha\beta}(k)=-4g^{\mu\nu}+\frac{2k_{\rho}}{k^{+}}\bigg[g^{\mu\rho}\gamma^{\nu}\gamma^{+}-g^{\rho\nu}\gamma^{\mu}\gamma^{+}+g^{\rho+}\gamma^{\mu}\gamma^{\nu}-g^{+\nu}\gamma^{\mu}\gamma^{\rho}+g^{+\mu}\gamma^{\nu}\gamma^{\rho}\bigg]
\end{equation}
and
\begin{equation}
\begin{split}
\gamma^{\alpha}\gamma^{\sigma}\gamma^{\mu}\gamma^{\nu}\gamma^{\beta}d_{\alpha\beta}(k)=\frac{2k_{\rho}}{k^{+}}&\bigg[g^{\rho+}\gamma^{\nu}\gamma^{\mu}\gamma^{\sigma}+g^{+\sigma}\gamma^{\mu}\gamma^{\nu}\gamma^{\rho}+g^{\rho\sigma}\gamma^{\mu}\gamma^{\nu}\gamma^{+}-g^{\rho\mu}\gamma^{\sigma}\gamma^{\nu}\gamma^{+}\\
&+g^{\rho\nu}\gamma^{\sigma}\gamma^{\mu}\gamma^{+}+g^{+\nu}\gamma^{\sigma}\gamma^{\mu}\gamma^{\rho}-g^{+\mu}\gamma^{\sigma}\gamma^{\nu}\gamma^{\rho}-g^{\rho+}\gamma^{\sigma}\gamma^{\mu}\gamma^{\nu}\bigg]
\end{split}
\end{equation}
which can easily be derived using the anticommutation relations of $\gamma$-matrices. Since there are no terms involving $k^{-}$ in the numerator, hence there are no arc contributions to the contour integral. Eq.(\ref{eq:LambdaMu3}) can thus be written as 
\begin{equation*}
\Lambda^{\mu}_{3_{p,p',q}}=ie^3\int \frac {d^{2}{\bf{k}}_{\perp}dk^{+}}{(2\pi)^4}\frac{\gamma^{\alpha}({k\llap/''_{on}}+m)\gamma^{\mu}\gamma^{+}\gamma^{\beta}}{2k^{+}2(p^{+}-k^{+})2(p'^{+}-k^{+})}\ I_{3}
\end{equation*}
where
\begin{equation*}
I_{3}=\int\frac{dk^{-}d_{\alpha\beta}(k)}{\bigg[k^{-}-\bigg[\frac{{\bf{k}}_{\perp}^{2}+\mu^{2}-i\epsilon}{2k^{+}}\bigg]\bigg]\bigg[p'^{-}-k^{-}-\bigg[\frac{({\bf{p}}'_{\perp}-{\bf{k}}_{\perp})^{2}+m^{2}-i\epsilon}{2(p'^{+}-k^{+})}\bigg]\bigg]}
\end{equation*}
which has poles at $k^{-}_{1}=\frac{{\bf{k}}_{\perp}^{2}+\mu^{2}-i\epsilon}{2k^{+}}$ and at $k^{-}_{2}=p'^{-}-\frac{({\bf{p}}'_{\perp}-{\bf{k}}_{\perp})^{2}+m^{2}-i\epsilon}{2(p'^{+}-k^{+})}$. For $k^{+}<0$, both $k^{-}_{1}$ and $k^{-}_{2}$ lie above while for $k^{+}>p'^{+}$, both lie below the real axis. Hence, $I_{3}=0$ for these ranges of $k^{+}$. Thus, $I_{3}$ is non-zero only for $0<k^{+}<p'^{+}$ and is equal, on closing the contour below the real axis, to the residue calculated at the pole $k^{-}_{1}$. Hence, 
\begin{equation*}
I_{3}=\frac{-2\pi i\theta(k^{+})\theta({p'}^{+}-k^{+})d_{\alpha\beta}(k_{on})}{\bigg[p'^{-}-\bigg[\frac{{\bf{k}}_{\perp}^{2}+\mu^{2}-i\epsilon}{2k^{+}}\bigg]-\bigg[\frac{({\bf{p}}'_{\perp}-{\bf{k}}_{\perp})^{2}+m^{2}-i\epsilon}{2(p'^{+}-k^{+})}\bigg]\bigg]}
\end{equation*}
Therefore, 
\begin{equation} \label{eq:vc_lambda3}
\Lambda^{\mu}_{3_{p,p',q}}=e^3\int \frac {d^{2}{\bf{k}}_{\perp}}{(4\pi)^3}\int_{0}^{p'^{+}}\frac{dk^{+}}{k^{+}k'^{+}k''^{+}}\frac{\gamma^{\alpha}({k\llap/''_{on}}+m)\gamma^{\mu}\gamma^{+}\gamma^{\beta}d_{\alpha\beta}(k_{on})}{(p'^{-}-k^{-}_{on}-k''^{-}_{on})}
\end{equation}
Similarly, the numerator of the integrand in $\Lambda^{\mu}_{4_{p,p',q}}$ of Eq.(\ref{eq:LambdaMu4}) can be written as
\begin{equation}
\begin{split}
\gamma^{\alpha}\gamma^{+}\gamma^{\mu}({k\llap/'_{on}}+m)\gamma^{\beta}d_{\alpha\beta}(k)=2\bigg[&k\llap/'_{on}\gamma^{\mu}\gamma^{+}+\gamma^{\mu}k\llap/'_{on}\gamma^{+}-g^{-\mu}\gamma^{+}k\llap/'_{on}\gamma^{+}+\frac{k^{\perp}{\cdot}g^{\perp\mu}\gamma^{+}k\llap/'_{on}\gamma^{+}}{k^{+}}\\
&+k'^{-}_{on}\gamma^{+}\gamma^{\mu}\gamma^{+}-\frac{k'^{\perp}{\cdot}k^{\perp}\gamma^{+}\gamma^{\mu}\gamma^{+}}{k^{+}}+2k'^{+}\gamma^{+}\gamma^{\mu}\gamma^{-}-\frac{k'^{+}\gamma^{+}\gamma^{\mu}\gamma^{\perp}{\cdot}k^{\perp}}{k^{+}}\\
&-g^{+\mu}\gamma^{+}k\llap/'_{on}\gamma^{-}+\frac{g^{+\mu}\gamma^{+}k\llap/'_{on}\gamma^{\perp}{\cdot}k^{\perp}}{k^{+}}-\gamma^{+}\gamma^{\mu}k\llap/'_{on}-mg^{+\mu}\gamma^{+}\gamma^{-}\\
&-\frac{mg^{+\mu}\gamma^{+}\gamma^{\perp}{\cdot}k^{\perp}}{k^{+}}\bigg]+4k^{-}\bigg[\frac{k'^{+}\gamma^{+}\gamma^{\mu}\gamma^{+}}{k^{+}}-\frac{g^{+\mu}\gamma^{+}k\llap/'_{on}\gamma^{+}}{k^{+}}\bigg]
\end{split}
\end{equation}
Here, the coefficient of $k^{-}$ is
\begin{equation*}
4\bigg[\frac{k'^{+}\gamma^{+}\gamma^{\mu}\gamma^{+}}{k^{+}}-\frac{g^{+\mu}\gamma^{+}k\llap/'_{on}\gamma^{+}}{k^{+}}\bigg]
\end{equation*}
and the two terms in the bracket cancel. Thus, in the case of $\Lambda^{\mu}_{4_{p,p',q}}$ too, there are no terms involving $k^{-}$ in the numerator. As a result, arc contributions to the contour integral are absent. Thus, Eq.(\ref{eq:LambdaMu4}) is written as
\begin{equation*}
\Lambda^{\mu}_{4_{p,p',q}}=ie^3\int \frac {d^{2}{\bf{k}}_{\perp}dk^{+}}{(2\pi)^4}\frac{\gamma^{\alpha}\gamma^{+}\gamma^{\mu}({k\llap/'_{on}}+m)\gamma^{\beta}}{2k^{+}2(p^{+}-k^{+})2(p'^{+}-k^{+})}\ I_{4}
\end{equation*}
where
\begin{equation*}
I_{4}=\int\frac{dk^{-}d_{\alpha\beta}(k)}{\bigg[k^{-}-\bigg[\frac{{\bf{k}}_{\perp}^{2}+\mu^{2}-i\epsilon}{2k^{+}}\bigg]\bigg]\bigg[p^{-}-k^{-}-\bigg[\frac{({\bf{p}}_{\perp}-{\bf{k}}_{\perp})^{2}+m^{2}-i\epsilon}{2(p^{+}-k^{+})}\bigg]\bigg]}
\end{equation*}
which has poles at $k^{-}_{1}=\frac{{\bf{k}}_{\perp}^{2}+\mu^{2}-i\epsilon}{2k^{+}}$ and at $k^{-}_{2}=p^{-}-\frac{({\bf{p}}_{\perp}-{\bf{k}}_{\perp})^{2}+m^{2}-i\epsilon}{2(p^{+}-k^{+})}$. Same arguments as for $I_{3}$ follow with $p'$ replaced by $p$. Hence, 
\begin{equation*}
I_{4}=\frac{-2\pi i\theta(k^{+})\theta({p}^{+}-k^{+})d_{\alpha\beta}(k_{on})}{\bigg[p^{-}-\bigg[\frac{{\bf{k}}_{\perp}^{2}+\mu^{2}-i\epsilon}{2k^{+}}\bigg]-\bigg[\frac{({\bf{p}}_{\perp}-{\bf{k}}_{\perp})^{2}+m^{2}-i\epsilon}{2(p^{+}-k^{+})}\bigg]\bigg]}
\end{equation*}
leading to 
\begin{equation} \label{eq:vc_lambda4}
\Lambda^{\mu}_{4_{p,p',q}}=e^3\int \frac {d^{2}{\bf{k}}_{\perp}}{(4\pi)^3}\int_{0}^{p^{+}}\frac{dk^{+}}{k^{+}k'^{+}k''^{+}}\frac{\gamma^{\alpha}\gamma^{+}\gamma^{\mu}({k\llap/'_{on}}+m)\gamma^{\beta}d_{\alpha\beta}(k_{on})}{(p^{-}-k^{-}_{on}-k'^{-}_{on})}
\end{equation}
Substituting Eqs.(\ref{eq:vc_lambda1}), (\ref{eq:vc_lambda2}), (\ref{eq:vc_lambda3}) and (\ref{eq:vc_lambda4}) in Eq.(\ref{eq:vc_split}), we see that
\begin{equation}\label{eq:vc_final}
\begin{split}
\Lambda^{\mu}_{p,p',q}=&e^3\int \frac {d^{2}{\bf{k}}_{\perp}}{(4\pi)^3}\int_{0}^{p'^{+}}\frac{dk^{+}}{k^{+}k'^{+}k''^{+}}\frac{\gamma^{\alpha}({k\llap/''_{on}}+m)\gamma^{\mu}({k\llap/'_{on}}+m)\gamma^{\beta}d_{\alpha\beta}(k_{on})}{(p^{-}-k^{-}_{on}-k'^{-}_{on})(p^{-}-q^{-}-k^{-}_{on}-k''^{-}_{on})}\\
-&e^3\int \frac {d^{2}{\bf{k}}_{\perp}}{(4\pi)^3}\int_{p'^{+}}^{p^{+}}\frac{dk^{+}}{k^{+}k'^{+}k''^{+}}\frac{\gamma^{\alpha}({k\llap/''_{on}}+m)\gamma^{\mu}({k\llap/'_{on}}+m)\gamma^{\beta}d_{\alpha\beta}(k_{on})}{(p^{-}-k^{-}_{on}-k'^{-}_{on})(p^{-}-p'^{-}-k'^{-}_{on}+k''^{-}_{on})}\\
&e^3\int \frac {d^{2}{\bf{k}}_{\perp}}{(4\pi)^3}\int_{0}^{p'^{+}}\frac{dk^{+}}{k^{+}k'^{+}k''^{+}}\frac{\gamma^{\alpha}({k\llap/''_{on}}+m)\gamma^{\mu}\gamma^{+}\gamma^{\beta}d_{\alpha\beta}(k_{on})}{(p'^{-}-k^{-}_{on}-k''^{-}_{on})}\\
&e^3\int \frac {d^{2}{\bf{k}}_{\perp}}{(4\pi)^3}\int_{0}^{p^{+}}\frac{dk^{+}}{k^{+}k'^{+}k''^{+}}\frac{\gamma^{\alpha}\gamma^{+}\gamma^{\mu}({k\llap/'_{on}}+m)\gamma^{\beta}d_{\alpha\beta}(k_{on})}{(p^{-}-k^{-}_{on}-k'^{-}_{on})}
\end{split}
\end{equation}

Taking into account  the normalization factor $\lambda$ given by $\lambda^{-1}=(2\pi)^{3/2}\sqrt{2p^{+}}\sqrt{2p'^{+}}\sqrt{2q^{+}}$, each term in Eq.(\ref{eq:vc_final}) is equal to the expressions for diagrams in Fig.(\ref{fig:reg_inst_photon})(a) through (d) viz. Eqs.(\ref{eq:vc_reg_diag1}), (\ref{eq:vc_reg_diag2}), (\ref{eq:vc_inst_ferm1}), (\ref{eq:vc_inst_ferm2}) respectively. The following inferences can thus be drawn:\\
(i) A general component of the vertex correction receives non-zero contributions from $\Lambda^{\mu}_{3_{p,p',q}}$ and $\Lambda^{\mu}_{4_{p,p',q}}$ which produce the instantaneous fermion diagrams. These contributions were absent in Ref.\citep{swati} and were not evaluated in Ref.\citep{mustaki} as only the `+' component of vertex correction was considered in both the works. These contributions arise from the off-shell part of fermion propagator.\\
(ii) The on-shell part of the fermion propagator when considered with the three-term photon propagator in the covariant theory corresponds to the regular vertex correction diagrams in LFTOPT.\\
(iii) Had we started with the two-term photon propagator in Eq.(\ref{eq:vc_main_3term}) instead of the three-term propagator used here,  Eq.(\ref{eq:LambdaMu2}) and in turn, Eq.(\ref{eq:vc_lambda2}) would  have been absent, thus retaining the last term in Eq.(\ref{eq:vc_lambda1}). This term corresponds to the instantaneous photon exchange diagram given in Fig.(\ref{fig:vertcorr_inst_photon}).\\
(iv) No contribution to vertex correction is received when we consider the off-shell parts of both propagators simultaneously.

\begin{figure}[h!]
\centering
\includegraphics[scale=0.55]{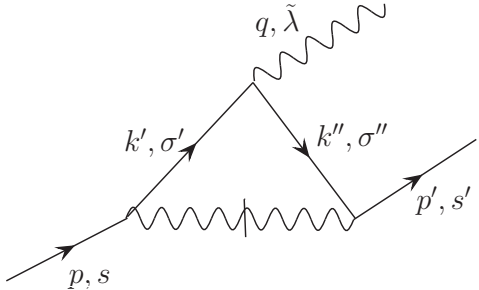}
\caption{Instantaneous photon exchange diagram}
\label{fig:vertcorr_inst_photon}
\end{figure}

\section{Summary}\label{sec:concl}

We have re-visited the issue of equivalence of covariant QED and LFQED with special emphasis on which form of the photon  propagator should be used in the proof of equivalence. We observe that in covariant formulation of QED, the three-term propagator is derived from the Lagrangian in Eq.(\ref{eq:equaltime_QED_Lag}) wherein both Lorentz condition as well as the gauge fixing condition $A^+ = 0 $ have been taken into account in the form of Lagrange's multiplier. In contrast, the LFQED Hamiltonian in Ref.\citep{mustaki}, which has been the reference point for the earlier work on this subject of equivalence at one loop level, is actually derived by eliminating the dependent degrees of freedom using only the LF gauge fixing condition. We, therefore, derive the LFQED Hamiltonian (Eq.(\ref{eq:LF_Ham})) following the procedure in Ref.\citep{mustaki} but  now also taking into account the Lorentz condition. We find that this Hamiltonian does not have the instantaneous photon exchange interaction and therefore the set of one loop graphs in this theory does not contain the diagrams involving instantaneous photon exchange. We consider this theory and show that indeed the one loop graphs of this theory can be obtained from the covariant expressions containing the three-term propagator by integrating over the light-front energy $k^-$. 
We compare our results with the work of Mantovani {\it et al.} who have established equivalence of one loop expressions with the expressions in Ref.\citep{mustaki} using the two-term photon propagator. Justification for using the two-term propagator, as given by Mantovani {\it et al.}, is that the contribution of the third term in the propagator cancels the contribution of the instantaneous  interaction (the last term in Eq.(\ref{eq:L_lf_int})) and therefore, it  is sufficient to work with the two-term propagator.  
Thus it is clear that the issue of equivalence as addressed by Matovani {\it et al.} and by us is at different levels. Our aim in this work is to establish the equivalence of covariant formulation of QED in LF gauge in instant form (Eq. (\ref{eq:equaltime_QED_Lag})) with the Hamiltonian formulation of LFQED in LF gauge  at one loop level.  Authors of Ref.\citep{mantovani}, on the other hand, have compared the Lagrangian formulation of LFQED in LF gauge (Eq.(\ref{eq:L_lf_int})) with the corresponding Hamiltonian version.
Since we start with the theory based on manifestly covariant Lagrangian in Eq.(\ref{eq:equaltime_QED_Lag}), the photon propagator will have the third term also and the LFQED Hamiltonian to be used for deriving LF Feynman rules for corresponding theory will be given by the Hamiltonian in Eq.(\ref{eq:LF_Ham}). On the other hand, if one starts with the interaction Lagrangian in Eq.(\ref{eq:L_lf_int}), there is no need to add the Lagrange's multiplier (since one has already used the condition to eliminate the unphysical degrees of freedom) and hence it is sufficient to use the two-term propagator only.

\par After clarifying the issue of the form of the photon propagator, we have established equivalence between the equal-time covariant QED and light-front time-ordered Hamiltonian QED at the level of one loop Feynman diagrams using two methods. In Sec. \ref{sec:review}, we used the method of splitting the photon propagator in on-shell and off-shell parts \citep{mantovani} to establish equivalence for fermion self-energy and vacuum polarization graphs. In Sec. \ref{sec:asymptotic}, we  introduced  an alternative method called the asymptotic method  and verified the results of Sec. \ref{sec:equi_self-en} using this method. In order to establish equivalence for a general component of one loop vertex correction, we first calculated the instantaneous fermion exchange graphs contributing to one loop vertex correction in Sec. \ref{sec:equi_vertcorr} which were not calculated in earlier works.  We have then extended our earlier proof of equivalence of vertex correction graphs to a general component of $\Lambda^\mu$.   
The asymptotic method was used by us in Ref.\citep{swati2} to show that the covariant expression for one loop vacuum polarization reproduces the corresponding LFQED diagrams on performing the $k^-$-integration.  
 In present work, we have shown that all the one loop self-energy and vertex correction diagrams of LFQED can also be reproduced starting from the covariant expressions using the asymptotic method. We establish this within both our approach as well as in the approach of Ref.\citep{mantovani}.
 

\section*{Acknowledgements}
AM would like to thank DST SERB for financial support under the grant no. EMR/2014/0000486 and International Centre for Theoretical Physics, Trieste, Italy for their kind hospitality. DB would like to acknowledge the financial support provided by University Grants Commission (UGC), India for carrying out this research and the travel support and hospitality provided by International Centre for Theoretical Physics (ICTP), Italy, where part of this work was carried out. 

\appendix 

\section{Basics and Conventions} \label{app:basics}
The 4-vector $x^{\mu}$, in LF coordinates, has the components $(x^{+}, x^{-}, {\bf{x}}^{\perp})$ where \\
$x^{+}=\frac{x^{0}+x^{3}}{\sqrt{2}}$, $x^{-}=\frac{x^{0}-x^{3}}{\sqrt{2}}$, ${\bf{x}}^{\perp}=(x^{1}, x^{2})$. \\
We use the following metric tensor:\\

$g_{\alpha\beta}=g^{\alpha\beta}=$
$ \begin{bmatrix}
0 & 1 & 0 & 0 \\
1 & 0 & 0 & 0 \\
0 & 0 & -1 & 0 \\
0 & 0 & 0 & -1 
\end{bmatrix} $\\

The following representation is used for the $\gamma$-matrices:
\begin{equation}
\gamma^{0}=\begin{bmatrix}
0 & I \\
I & 0 \\
\end{bmatrix}, \gamma^{k}=\begin{bmatrix}
0 & -\sigma_{k} \\
\sigma_{k} & 0 \\
\end{bmatrix}, \gamma^{+}=\frac{\gamma^{0}+\gamma^{3}}{\sqrt{2}}, \gamma^{-}=\frac{\gamma^{0}-\gamma^{3}}{\sqrt{2}}
\end{equation}

The $\gamma$-matrices satisfy
\begin{equation}
 \begin{split}
 &\{\gamma^{\alpha}, \gamma^{\beta}\}=2g^{\alpha\beta}\\ &(\gamma^{+})^{2}=(\gamma^{-})^{2}=0\\
 &(\gamma^{0})^{\dagger}=\gamma^{0}\\
 &(\gamma^{k})^{\dagger}=-\gamma^{k} for\ k=1,2,3\\
 &\gamma^{+}\gamma^{-}\gamma^{+}=2\gamma^{+}, \gamma^{-}\gamma^{+}\gamma^{-}=2\gamma^{-}\\ &\gamma^{\alpha}\gamma^{\mu}\gamma^{\beta}d_{\alpha\beta}(k)=\frac{2}{k^{+}}(\gamma^{+}k^{\mu}+g^{+\mu}k\llap/)
 \end{split}
\end{equation}
The Dirac spinors satisfy  the following properties:
\begin{equation}
 \begin{split}
 &\bar{u}_{p,s}u_{p,s'}=-\bar{v}_{p,s}v_{p,s'}=2m\delta_{ss'}\\
 &\bar{u}_{p,s}\gamma^{\mu}u_{p,s'}=-\bar{v}_{p,s}\gamma^{\mu}v_{p,s'}=2p^{\mu}\delta_{ss'}
 \end{split}
\end{equation}
and the completeness relations
\begin{equation}
 \begin{split}
 &\sum_{s=\pm1/2}u_{p,s}\bar{u}_{p,s}=p\llap/+m\\ &\sum_{s=\pm1/2}v_{p,s}\bar{v}_{p,s}=p\llap/-m
 \end{split}
\end{equation}
For photon polarizations, we choose
\begin{equation}
\epsilon^{1}_{\mu}=\bigg(\frac{p^{1}}{p^{+}},0,-1,0\bigg), \epsilon^{2}_{\mu}=\bigg(\frac{p^{2}}{p^{+}},0,0,-1\bigg)
\end{equation}
The null-plane Hamiltonian is
\begin{equation*}
P^{-}=H_{0}+V_{1}+V_{2}+V_{3}
\end{equation*}
when the gauge field satisfies the LF gauge condition and
\begin{equation*}
P^{-}=H_{0}+V_{1}+V_{2}
\end{equation*}
when the gauge field satisfies the LF gauge condition as well as the Lorentz condition.\\
Here, in addition to the free Hamiltonian $H_{0}$ and the standard three-point order-$e$ interaction
\begin{equation} \label{eq:V1_expression}
V_1=e\int d^{2}{\bf{x}}_{\perp}dx^{-}\xi\gamma^{\mu}\xi a_{\mu},
\end{equation}
there exist additional order-$e^{2}$ non-local interactions
\begin{equation} \label{eq:V2_expression}
V_{2}=-\frac{i}{4}e^{2}\int d^{2}{\bf{x}}_{\perp}dx^{-}dy^{-}\epsilon(x^{-}-y^{-})(\bar{\xi}a_{k}\gamma^{k})(x)\gamma^{+}(a_{j}\gamma^{j}\xi)(y)
\end{equation}
corresponding to an instantaneous fermion exchange and
\begin{equation}
V_{3}=-\frac{e^{2}}{4}\int d^{2}{\bf{x}}_{\perp}dx^{-}dy^{-}(\bar{\xi}\gamma^{+}\xi)(x)|x^{-}-y^{-}|(\bar{\xi}\gamma^{+}\xi)(y)
\end{equation}
corresponding to an instantaneous photon exchange.

\section{} \label{app:details_sec_III}
\subsection{LFTOPT Diagram Calculations for Vertex Correction}

In this appendix, we present the details of the calculation of the expression for the diagram of Fig.\ref{fig:reg_inst_photon}(c). The transition amplitude that contributes to one loop correction arising from Fig.\ref{fig:reg_inst_photon}(c) is
\begin{equation}\label{eq:vc_with_matrix_elements}
 \begin{split}
 T^{(c)}_{p,p',q} = &\mel{p',s';q,\tilde{\lambda}}{V_1\frac{1}{p^{-}-H_0}V_2}{p,s}\\
 =&\int_{-\infty}^{+\infty}d^{2}{\bf{k}}''_{\perp}d^{2}{\bf{k}}_{\perp}d^{2}{\bf{k}}''_{1\perp}d^{2}{\bf{k}}_{1\perp}\int_{0}^{\infty}dk''^{+}dk^{+}dk''^{+}_{1}dk^{+}_{1}\sum_{\sigma'',\lambda,\sigma''_1,\lambda_1}\mel{p',s';q,\tilde{\lambda}}{V_1}{k'',\sigma'';k,\lambda;q,\tilde{\lambda}}\\
 &\mel{k'',\sigma'';k,\lambda;q,\tilde{\lambda}}{\frac{1}{p^{-}-H_0}}{k''_1,\sigma''_1;k_1,\lambda_1;q,\tilde{\lambda}}\mel{k''_1,\sigma''_1;k_1,\lambda_1;q,\tilde{\lambda}}{V_2}{p,s}\\
 =&\int\frac{d^{3}{\bf{k}}''d^{3}{\bf{k}}d^{3}{\bf{k}}''_1d^{3}{\bf{k}}_1\theta(k''^+)\theta(k^+)\theta(k''^{+}_1)\theta(k^{+}_1)}{p^{-}-k''^{-}_{1}-k^{-}_{1}-q^{-}}\sum_{\sigma'',\lambda,\sigma''_1,\lambda_1}\mel{p',s';q,\tilde{\lambda}}{V_1}{k'',\sigma'';k,\lambda;q,\tilde{\lambda}}\\
 &\braket{k'',\sigma'';k,\lambda;q,\tilde{\lambda}}{k''_1,\sigma''_1;k_1,\lambda_1;q,\tilde{\lambda}}\mel{k''_1,\sigma''_1;k_1,\lambda_1;q,\tilde{\lambda}}{V_2}{p,s}\\
 =&\int\frac{d^{3}{\bf{k}}''d^{3}{\bf{k}}\theta(k''^+)\theta(k^+)}{p^{-}-k''^{-}-k^{-}-q^{-}}\sum_{\sigma'',\lambda}\mel{p',s';q,\tilde{\lambda}}{V_1}{k'',\sigma'';k,\lambda;q,\tilde{\lambda}}\mel{k'',\sigma'';k,\lambda;q,\tilde{\lambda}}{V_2}{p,s}
 \end{split}
\end{equation}
where the orthonormality of states is used to arrive at the final step. Using Eqns.(\ref{eq:V1_expression}) and (\ref{eq:V2_expression}), the matrix elements in the above expression for transition amplitude, on Fourier expanding the fields, are written as:
\begin{equation*}
 \begin{split}
 \mel{p',s';q,\tilde{\lambda}}{V_1}{k'',\sigma'';k,\lambda;q,\tilde{\lambda}} =&e\int d^{2}{\bf{x}}_{\perp}dx^{-}\int_{-\infty}^{+\infty}\frac{d^{2}{\bf{p}}_{1\perp}d^{2}{\bf{p}}_{2\perp}d^{2}{\bf{q}}_{1\perp}}{(2\pi)^{9/2}\sqrt{8}}\int_{0}^{\infty}\frac{dp_{1}^{+}dp_{2}^{+}dq_{1}^{+}}{\sqrt{p_{1}^{+}p_{2}^{+}q_{1}^{+}}}\sum_{s_{1},s_{2},\lambda_{1}}\\
 &\bra{p',s';q,\tilde{\lambda}}{\big[\bar{u}_{p_{1},s_{1}}e^{ip_{1}\boldsymbol{\cdot}x}b^{\dagger}_{p_{1},s_{1},x}+\bar{v}_{p_{1},s_{1}}e^{-ip_{1}\boldsymbol{\cdot}x}d_{p_{1},s_{1},x}\big]\gamma^\mu\\
 &\big[u_{p_{2},s_{2}}e^{-ip_{2}\boldsymbol{\cdot}x}b_{p_{2},s_{2},x}+v_{p_{2},s_{2}}e^{ip_{2}\boldsymbol{\cdot}x}d^{\dagger}_{p_{2},s_{2},x}\big]{\epsilon_{\mu}^{\lambda_{1}}}(q_1)\\
 &[e^{-iq_{1} \boldsymbol{\cdot} x}a_{q_{1},\lambda_{1},x}+e^{iq_{1} \boldsymbol{\cdot} x}a^{\dagger}_{q_{1},\lambda_{1},x}]}\ket{k'',\sigma'';k,\lambda;q,\tilde{\lambda}}
 \end{split}
 \end{equation*}
where $e^{ip_{1}\boldsymbol{\cdot}x} = e^{i[p_{1}^{+}x^{-}-{\bf{p}}_{1\perp}\boldsymbol{\cdot}{\bf{x}}_\perp]}$ etc.\\
Using $\bra{p',s';q,\tilde{\lambda}}b^{\dagger}_{p_{1},s_{1},x}b_{p_{2},s_{2},x}a_{q_{1},\lambda_{1},x}\ket{k'',\sigma'';k,\lambda;q,\tilde{\lambda}}=\delta^{3}({\bf{q}}_{1}-{\bf{k}})\delta_{\lambda_{1}\lambda}\delta^{3}({\bf{p}}_{2}-{\bf{k}}'')\delta_{s_{2}\sigma''}\\
 \delta^{3}({\bf{p}}_{1}-{\bf{p}}')\delta_{s_{1}s'}$\\
 where $\delta^{3}({\bf{q}}_{1}-{\bf{k}}) = \delta^{2}({\bf{q}}_{1\perp}-{\bf{k}}_{\perp})\delta(q_{1}^{+}-k^{+})$ etc.,\\
 we obtain
\begin{equation}\label{eq:matrix_element_V1}
  \begin{split}
  \mel{p',s';q,\tilde{\lambda}}{V_1}{k'',\sigma'';k,\lambda;q,\tilde{\lambda}}
  = & e\int d^{2}{\bf{x}}_{\perp}dx^{-}\int_{-\infty}^{+\infty}\frac{d^{3}{\bf{p}}_{1}d^{3}{\bf{p}}_{2}d^{3}{\bf{q}}_{1}\theta(p_{1}^{+})\theta(p_{2}^{+})\theta(q_{1}^{+})}{(2\pi)^{9/2}\sqrt{8}{\sqrt{p_{1}^{+}p_{2}^{+}q_{1}^{+}}}}\sum_{s_{1},s_{2},\lambda_{1}}\\
  &\bar{u}_{p_{1},s_{1}}e^{ip_{1}\boldsymbol{\cdot}x}\gamma^{\mu}u_{p_{2},s_{2}}e^{-ip_{2}\boldsymbol{\cdot}x}{\epsilon_{\mu}^{\lambda_{1}}}(q_1)e^{-iq_{1}\boldsymbol{\cdot}x}\\
  &\delta^{3}({\bf{q}}_{1}-{\bf{k}})\delta_{\lambda_{1}\lambda}\delta^{3}({\bf{p}}_{2}-{\bf{k}}'')\delta_{s_{2}\sigma''}\delta^{3}({\bf{p}}_{1}-{\bf{p}}')\delta_{s_{1}s'}\\
  =& e\int\frac{d^{2}{\bf{x}}_{\perp}dx^{-}\theta(p'^{+})\theta(k''^{+})\theta(k^{+})}{(2\pi)^{9/2}\sqrt{8}{\sqrt{p'^{+}k''^{+}k^{+}}}}\bar{u}_{p',s'}\gamma^{\mu}u_{k'',\sigma''}{\epsilon_{\mu}^{\lambda}(k)}e^{i(p'-k''-k)\boldsymbol{\cdot}x}\\
  =& \frac{e}{(2\pi)^{3/2}}\frac{1}{\sqrt{8}}\frac{1}{\sqrt{p'^{+}k''^{+}k^{+}}}\bar{u}_{p',s'}\gamma^{\mu}u_{k'',\sigma''}{\epsilon_{\mu}^{\lambda}(k)}\\
  &\delta^{3}[{\bf{k}}''-({\bf{p}}'-{\bf{k}})]\theta(p'^{+})\theta(k''^{+})\theta(k^{+})
  \end{split}
\end{equation}
Similarly,
\begin{equation*}
 \begin{split}
 \mel{k'',\sigma'';k,\lambda;q,\tilde{\lambda}}{V_2}{p,s} =& \frac{-ie^2}{4}\int d^{2}{\bf{y}}_{\perp}dy^{-}dz^{-}\epsilon(y^{-}-z^{-})\int_{-\infty}^{+\infty}\frac{d^{2}{\bf{p}}_{3\perp}d^{2}{\bf{p}}_{4\perp}d^{2}{\bf{q}}_{2\perp}d^{2}{\bf{q}}_{3\perp}}{(2\pi)^{6}\ 4}\\
 &\int_{0}^{\infty}\frac{dp_{3}^{+}dp_{4}^{+}dq_{2}^{+}dq_{3}^{+}}{\sqrt{p_{3}^{+}p_{4}^{+}q_{2}^{+}q_{3}^{+}}}\sum_{s_{3},s_{4},\lambda_{2},\lambda_{3}}\bra{k'',\sigma'';k,\lambda;q,\tilde{\lambda}}{\big[\bar{u}_{p_{3},s_{3}}e^{ip_{3}\boldsymbol{\cdot}y}b^{\dagger}_{p_{3},s_{3},y}+\\
 &\bar{v}_{p_{3},s_{3}}e^{-ip_{3}\boldsymbol{\cdot}y}d_{p_{3},s_{3},y}\big]{\epsilon_{k}^{\lambda_{2}}}(q_2)[e^{-iq_{2}\boldsymbol{\cdot}y}a_{q_{2},\lambda_{2},y}+e^{iq_{2}\boldsymbol{\cdot}y}a^{\dagger}_{q_{2},\lambda_{2},y}]\gamma^{k}\gamma^{+}\gamma^{j}{\epsilon_{j}^{\lambda_{3}}}(q_3)\\
 &\big[e^{-iq_{3}\boldsymbol{\cdot}z}a_{q_{3},\lambda_{3},z}+e^{iq_{3}\boldsymbol{\cdot}z}a^{\dagger}_{q_{3},\lambda_{3},z}\big]\big[u_{p_{4},s_{4}}e^{-ip_{4}\boldsymbol{\cdot}z}b_{p_{4},s_{4},z}+v_{p_{4},s_{4}}e^{ip_{4}\boldsymbol{\cdot}z}d^{\dagger}_{p_{4},s_{4},z}\big]\ket {p,s}}
 \end{split}
\end{equation*}
Again using $\bra{k'',\sigma'';k,\lambda;q,\tilde{\lambda}}b^{\dagger}_{p_{3},s_{3},y}a^{\dagger}_{q_{2},\lambda_{2},y}a^{\dagger}_{q_{3},\lambda_{3},z}b_{p_{4},s_{4},z}\ket{p,s}=\delta^{3}({\bf{p}}_{4}-{\bf{p}})\delta_{s_{4}s}\delta^{3}({\bf{q}}_{3}-{\bf{k}})\delta_{\lambda_{3}\lambda}\\
 \delta^{3}({\bf{q}}_{2}-{\bf{q}})\delta_{\lambda_{2}\tilde{\lambda}}\delta^{3}({\bf{p}}_{3}-{\bf{k}}'')\delta_{s_{3}\sigma''}$,\\
we obtain
 \begin{equation}\label{eq:matrix_element_V2}
   \begin{split}
   \mel{k'',\sigma'';k,\lambda;q,\tilde{\lambda}}{V_2}{p,s} =& \frac{-ie^2}{4}\int d^{2}{\bf{y}}_{\perp}dy^{-}dz^{-}\int_{-\infty}^{+\infty} \frac{d^{3}{\bf{p}}_{3}d^{3}{\bf{p}}_{4}d^{3}{\bf{q}}_{2}d^{3}{\bf{q}}_{3}\theta(p_{3}^{+})\theta(p_{4}^{+})\theta(q_{2}^{+})\theta(q_{3}^{+})}{(2\pi)^{6}\ 4\ {\sqrt{p_{3}^{+}p_{4}^{+}q_{2}^{+}q_{3}^{+}}}}\\
   &\epsilon(y^{-}-z^{-})\sum_{s_{3},s_{4},\lambda_{2},\lambda_{3}}\bar{u}_{p_{3},s_{3}}e^{ip_{3}\boldsymbol{\cdot}y}{\epsilon_{k}^{\lambda_{2}}}(q_2)e^{iq_{2}\boldsymbol{\cdot}y}\gamma^{k}\gamma^{+}\gamma^{j}{\epsilon_{j}^{\lambda_{3}}}(q_3)e^{iq_{3}\boldsymbol{\cdot}z}u_{p_{4},s_{4}}e^{-ip_{4}\boldsymbol{\cdot}z}\\
   &\delta^{3}({\bf{p}}_{4}-{\bf{p}})\delta_{s_{4}s}\delta^{3}({\bf{q}}_{3}-{\bf{k}})\delta_{\lambda_{3}\lambda}\delta^{3}({\bf{q}}_{2}-{\bf{q}})\delta_{\lambda_{2}\tilde{\lambda}}\delta^{3}({\bf{p}}_{3}-{\bf{k}}'')\delta_{s_{3}\sigma''}\\
   =&\frac{-ie^2}{4}\int\frac{d^{2}{\bf{y}}_{\perp}dy^{-}dz^{-}\theta(k''^{+})\theta(q^{+})\theta(k^{+})\theta(p^{+})}{(2\pi)^{6}\ 4\ {\sqrt{k''^{+}q^{+}k^{+}p^{+}}}}\epsilon(y^{-}-z^{-})\bar{u}_{k'',\sigma''}e^{ik''\boldsymbol{\cdot}y}{\epsilon_{k}^{\tilde{\lambda}}}(q)\\
   & e^{iq\boldsymbol{\cdot}y}\gamma^{k}\gamma^{+}\gamma^{j}{\epsilon_{j}^{\lambda}}(k)e^{ik\boldsymbol{\cdot}z}u_{p,s}e^{-ip\boldsymbol{\cdot}z}\\
   =&\frac{-e^2}{8}\int\frac{d^{2}{\bf{y}}_{\perp}dy^{-}\theta(k''^{+})\theta(q^{+})\theta(k^{+})\theta(p^{+})}{(2\pi)^{6}\ {\sqrt{k''^{+}q^{+}k^{+}p^{+}}}\ (k^{+}-p^{+})}\bar{u}_{k'',\sigma''}\gamma^{k}\gamma^{+}\gamma^{j}u_{p,s}\\
   &{\epsilon_{j}^{\lambda}}(k){\epsilon_{k}^{\tilde{\lambda}}}(q)e^{i(k''+q+k-p)\boldsymbol{\cdot}y}\\
   =&\frac{e^2}{(2\pi)^{3}}\frac{1}{8}\frac{\theta(k''^{+})\theta(q^{+})\theta(k^{+})\theta(p^{+})}{{\sqrt{k''^{+}q^{+}k^{+}p^{+}}}\ (p^{+}-k^{+})}\bar{u}_{k'',\sigma''}\gamma^{k}\gamma^{+}\gamma^{j}u_{p,s}{\epsilon_{j}^{\lambda}}(k){\epsilon_{k}^{\tilde{\lambda}}}(q)\\
   &\delta^{3}({\bf{k}}''+{\bf{q}}+{\bf{k}}-{\bf{p}})
   \end{split}
  \end{equation}
where the identity
\begin{equation*}
\int dz^{-}f(z^{-})\epsilon(y^{-}-z^{-})=\frac{2}{\partial_{-}}f(y)
\end{equation*}
is used for arriving at the above result. Substituting Eqns.(\ref{eq:matrix_element_V1}) and (\ref{eq:matrix_element_V2}) in Eq.(\ref{eq:vc_with_matrix_elements}), we get,
\begin{equation*}
 \begin{split}
 T^{(c)}_{p,p',q} = & e^{3}\lambda\int\frac{d^{3}{\bf{k}}''d^{3}{\bf{k}}\theta(k''^+)\theta(k^+)\theta(p^+)\theta(p'^+)\theta(q^+)}{(4\pi)^{3}k^{+}k''^{+}(p^{+}-k^{+})}\sum_{\sigma'',\lambda}\frac{\bar{u}_{p',s'}\gamma^{\mu}u_{k'',\sigma''}\bar{u}_{k'',\sigma''}\gamma^{k}\gamma^{+}\gamma^{j}u_{p,s}}{(p^{-}-k''^{-}-k^{-}-q^{-})}\\
 &{\epsilon_{\mu}^{\lambda}(k)}{\epsilon_{j}^{\lambda}}(k){\epsilon_{k}^{\tilde{\lambda}}}(q)\delta^{3}({\bf{p}}'-{\bf{k}}-{\bf{k}}'')\delta^{3}({\bf{k}}''+{\bf{q}}+{\bf{k}}-{\bf{p}})
 \end{split}
\end{equation*}
where $\lambda=\frac{1}{(2\pi)^{3/2}\sqrt{2p^{+}}\sqrt{2p'^{+}}\sqrt{2q^{+}}}$.\\
Using the completeness relations
\begin{equation*}
\sum_{s=\pm1/2}u_{p,s}\bar{u}_{p,s}=p\llap/+m
\end{equation*}
and
\begin{equation*}
\sum_{\lambda=1,2}\epsilon_{\mu}^{\lambda}(p)\epsilon_{\nu}^{\lambda}(p)=d_{\mu\nu}(p)=-g_{\mu\nu}+\frac{\delta_{{\mu}+}p_{\nu}+\delta_{{\nu}+}p_{\mu}}{p^+}
\end{equation*}
and performing the $k''$-integral using the delta functions, the amplitude for the diagram in Fig.\ref{fig:reg_inst_photon}(c) reduces to Eq.(\ref {eq:amp1}).\\
A similar calculation leads to Eq.(\ref{eq:amp2}) for the amplitude of Fig.\ref{fig:reg_inst_photon}(d).

\subsection{Calculation of numerators of $T^{(c)}_{p,p',q}$ and $T^{(d)}_{p,p',q}$}

Here we present the steps used for simplifying the numerator of Eqs.(\ref{eq:amp1}) and (\ref{eq:amp2}) to obtain Eqs.(\ref{eq:vc_amp_2a}) and (\ref{eq:vc_amp_2b}) respectively.\\
First we observe that since $(\gamma^{+})^{2}=0,$ and $\epsilon_{-}=0$, the `+' and `-' components of $\mu$ do not contribute.\\
Next we consider
\begin{equation*}
\begin{split}
 &\gamma^{\alpha}({k\llap/}''+m)\gamma^{\mu}\gamma^{+}\gamma^{\beta}d_{\alpha\beta}(k)\epsilon^{\tilde{\lambda}}_{\mu}(q)\\
 =&\gamma^{\alpha}({k\llap/}''+m)\gamma^{k}\gamma^{+}\gamma^{\beta}d_{\alpha\beta}(k)\epsilon^{\tilde{\lambda}}_{k}(q)\\
 =& [\gamma^{\alpha}({k\llap/}''+m)\gamma^{k}\gamma^{+}\gamma^{-}d_{\alpha-}(k)+\gamma^{\alpha}({k\llap/}''+m)\gamma^{k}\gamma^{+}\gamma^{j}d_{\alpha j}(k)]\epsilon^{\tilde{\lambda}}_{k}(q)\\
\end{split}
\end{equation*}
Now,
\begin{equation*}
\begin{split}
\gamma^{\alpha}({k\llap/}''+m)\gamma^{k}\gamma^{+}\gamma^{-}d_{\alpha-}(k)=& \gamma^{\alpha}({k\llap/}''+m)\gamma^{k}\gamma^{+}\gamma^{-}\bigg[-g_{\alpha-}+\frac{\delta_{\alpha+}k_{-}+\delta_{-+}k_{\alpha}}{k^{+}}\bigg]\\
 =& -\gamma^{+}({k\llap/}''+m)\gamma^{k}\gamma^{+}\gamma^{-}+\gamma^{+}({k\llap/}''+m)\gamma^{k}\gamma^{+}\gamma^{-}\bigg(\frac {k_{-}}{k^{+}}\bigg)\\
 =& \ 0
\end{split}
\end{equation*}
Therefore,
\begin{equation} \label{eq:appendix_b4}
\gamma^{\alpha}({k\llap/}''+m)\gamma^{\mu}\gamma^{+}\gamma^{\beta}d_{\alpha\beta}(k)\epsilon^{\tilde{\lambda}}_{\mu}(q)=\gamma^{\alpha}({k\llap/}''+m)\gamma^{k}\gamma^{+}\gamma^{j}d_{\alpha j}(k)\epsilon^{\tilde{\lambda}}_{k}(q)
\end{equation}
Similarly,
\begin{equation*}
\begin{split}
 &\gamma^{\alpha}\gamma^{+}\gamma^{\mu}({k\llap/}'+m)\gamma^{\beta}d_{\alpha\beta}(k)\epsilon^{\tilde{\lambda}}_{\mu}(q)\\
 =&\gamma^{\alpha}\gamma^{+}\gamma^{j}({k\llap/}'+m)\gamma^{\beta}d_{\alpha\beta}(k)\epsilon^{\tilde{\lambda}}_{j}(q)\\
 =& [\gamma^{-}\gamma^{+}\gamma^{j}({k\llap/}'+m)\gamma^{\beta}d_{-\beta}(k)+\gamma^{k}\gamma^{+}\gamma^{j}({k\llap/}'+m)\gamma^{\beta}d_{k\beta}(k)]\epsilon^{\tilde{\lambda}}_{j}(q)\\
\end{split}
\end{equation*}
Using
\begin{equation*}
\begin{split}
\gamma^{-}\gamma^{+}\gamma^{j}({k\llap/}'+m)\gamma^{\beta}d_{-\beta}(k)=& \gamma^{-}\gamma^{+}\gamma^{j}({k\llap/}'+m)\gamma^{\beta}\bigg[-g_{-\beta}+\frac{\delta_{-+}k_{\beta}+\delta_{\beta+}k_{-}}{k^{+}}\bigg]\\
 =& -\gamma^{-}\gamma^{+}\gamma^{j}({k\llap/}'+m)\gamma^{+}+\gamma^{-}\gamma^{+}\gamma^{j}({k\llap/}'+m)\gamma^{+}\bigg(\frac {k_{-}}{k^{+}}\bigg)\\
 =& \ 0
 \end{split}
\end{equation*}
we obtain
\begin{equation} \label{eq:appendix_b5}
\gamma^{\alpha}\gamma^{+}\gamma^{\mu}({k\llap/}'+m)\gamma^{\beta}d_{\alpha\beta}(k)\epsilon^{\tilde{\lambda}}_{\mu}(q)=\gamma^{k}\gamma^{+}\gamma^{j}({k\llap/}'+m)\gamma^{\beta}d_{k\beta}(k)\epsilon^{\tilde{\lambda}}_{j}(q)
\end{equation}

\bibliography{sample}

\end{document}